\documentclass[11pt,a4paper]{article}
\usepackage[utf8]{inputenc}
\usepackage{color}
\usepackage{xcolor}
\usepackage{amsmath}
\usepackage{amssymb}
\usepackage{amsfonts}
\usepackage{float}
\usepackage[header, toc]{appendix}
\usepackage{tabularx,colortbl}
\usepackage{enumerate}
\usepackage{amsmath}
\usepackage{tikz}
\usepackage{lscape}
\usepackage{cancel}
\usepackage{calrsfs}
\usepackage{amsmath}
    \numberwithin{equation}{section}
    \numberwithin{table}{section}
    \numberwithin{figure}{section}

\newcommand{\be}{\begin{equation}}
\newcommand{\ee}{\end{equation}}
\newcommand{\ba}{\begin{eqnarray}}
\newcommand{\ea}{\end{eqnarray}}
\newcommand{\p}{\partial}
\newcommand{\lag}{ \mathcal{L}}
\newcommand{\lagK}{ \mathcal{L}_K}

\newcommand{\lagGF}{ \mathcal{L}_{GF}}

\def\no{\nonumber}

\usepackage{amssymb}
\usepackage{hyperref}

\begin{document}
\begin{center}
    {\Large \bf Abelian Constructivist Lagrangian}\\	
    {\large $^\dagger$J. Chauca \footnote{jchauca@gmail.com}},
    {\large $^\dagger$R. Doria\footnote{doria@aprendanet.com.br} and}
    {\large $^\dagger$L.S. Mendes \footnote{santiago.petropolis@outlook.com}}
    \\[0.5cm]
    
    {\large $^\dagger$Aprendanet, Petropolis, Brazil; Quarks, Petropolis, Brazil}\\
\end{center}

\begin{abstract}
    Constructivist lagrangian propiates a diverse approach to field theory. Introduce the set action. Consider fields families under a same symmetry group. The resulting fields set extends the standard atomist field theory to a whole field theory.
    
    An associative physics is proposed. The grouping physics. The relationship between the part and the whole is considered. A third quantum type beyond Planck granularity and quantum mechanics wave-particle is obtained. The quantum inserted in the whole. Differentiated energy packets are formed. A quantum system is constituted.
     
    The abelian grouping physics is considered. The simplest whole unity. Set action in terms of $U(1)$ symmetry. The correspondent constructivist lagrangian is studied. A new type of individuation called whole quantum is derived. An abelian quantum system is constituted. The usual atomist gauge symmetry is preserved, but, constructivist properties are generated.

    Quantum system with own norm is a necessary theoretical argument. More is different, once time said P.W. Anderson. Physics is challenged to make the passage from an isolated particle to a quantum system. A theory to describe the physics of part in the whole. A challenge to be interpreted under gauge symmetry. A symmetry of difference is proposed. Quantum diversities enlarging the meanings of interaction, induction, connectivity. 
    
    A quantum system is introduced. Its elementarity is identified as a third quantum type. It is called whole quantum or variety. This quantum of a many particles system appears with a new physicality. The corresponding set action derives antireductionist physical laws under gauge symmetry. Ruled by associativity, set transformation, evolution. Associativitity providing quantum under set, diversity, interdependence, nonlinearity, chance. Set transformations performing a whole determinism with directive conducted by gauge parameter and circumstances under lagrangian free coefficients. Generating a set physics with growth, evolution, emergence, complexity. An evolving quantum transforming their quantum numbers. 
    
    The abelian constructivist lagrangian is explored. A quantum system assembled by fields families and gauge scalars is performed. It arises an environmental physics. Gauge scalars form substrutures with realities and potentialities. The volume of circumstances for each gauge scalar is calculated. Nonvirtual relationships are derived. Physical entities as masses, charges, coupling constants are expressed under constructivist properties. Functionalities will lead them to a physical behaviour beyond four interactions.
\end{abstract}

\section{Introduction}

${}$\indent
Quantum theory is a definitive step on the path of knowledgement about nature. It fixes energy in small amounts. An history which may be divided in three parts. The classical quantum theory started with Planck in 1900, electron, photoelectic effect, Rutherford-Bohr atom, proton, and in 1923 concluded with Compton words “... a radiation quantum carries with its momentum as well as energy.” [1]. 

The second phase, initiated in 1924 with du Broglie wave-particle, 1926 through quantum mechanics, 1927 with Heisenberg uncertainity principle, Dirac electromagnetic quantization, Jordan and Born relativistic wave equations [2]. They introduced quantum mechanics at old quantum theory. And in 1928, the quanta theory was related with relativity through Dirac equation [3]. Since then, physics has been developing an atomist quantum field theory incorporating granularity, wave-particle and relativity. As consequence, diverse theories have been developed as QED, QCD, electroweak, grandunification, supersymmetry, supergravity, extra dimension, superstrings, M-theory [4].

Nevertheless, a principle is missing in the quantum idea. The set notion. As thermodynamics considers that universe events tend to entropy, the principle of totality emphasizes the manifestation to association. Atoms, molecules,...galaxies are grasping how the universe organizes different forms of conglomerates. A perspective where the set would be the constitutive atom of the universe.

At current physics prevaleces the reductionist cognition. Parts derived as subunits summatory. However, the principles of associativity, confinement, complexity lead physics to the set action. There is a quantum physics under the whole unity to be understood. A performance beyond ultimate constituents. Consider the set as the fundamental generator and the correspondent part inserted in the whole. A quantum physics with a whole-part duality is generated. As consequence, a new physical elementarity is introduced. Parts become varieties of the whole environment. Variety defined as the whole quantum inserted in a quantum system.

Differents origins coming from Kaluza-Klein, fibre bundle, supersymmetry, non-linear $\sigma$-models, CP-models, degrees of freedom counting are showing the presence of fields set $\{A_{\mu I}\}$ encoded in an abelian group [5]. These various demonstrations are supporting the totality principle in gauge theory. Allowing, mathematically, the set notion and corresponding whole quantum be stipulated through gauge symmetry.

A third quantum physics is expressed. There is a whole quantum field theory beyond ultimate constituents to be constituted. An integral reality to be taken. Quantum derived from the whole association. A grouping physics crossing the Newton mechanics reductionism and quantum field theory atomism. A physics of variety originated from the relationship between ultimate constituents and set action.

Constructivist lagrangian turns out the formalism to prescribe the set action and corresponding third quantum. Quantum in a whole enlarges systemically the interaction, induction, connectivity. Antireductionist properties are established. A physics under set, diversity, interdependence, non-linearity, chance, directive and circumstance, growth, evolution, emergence, complexity. Features producing functionalities beyond four interactions.

Thus, the research will be about the whole abelian quantum physics.  Consider an abelian lagrangian enlarged by the antireductionist gauge symmetry. A symmetry identified first under a fields set $\{A_{\mu I}\}$ associated by a common gauge parameter; rewritten, in terms of a constructor basis $\{D_{\mu}, X_{\mu}^{i}\}$; subsequently, do physics with a realistic quanta set. Consider the constructivist lagrangian as exposed at Appendice A in terms of a physical basis $\{G_{\mu I}\}$. On it, analyze the symmetry of difference features.

\section{Fields set $\{G_{\mu I}\}$}
${}$\indent
 
 Considering the physical principle where nature acts through groupings, the first purpose is to study an antireductionist gauge theory (AGT). Introduce an abelian set of N-vector fields interconnected under a common gauge parameter. The significance is that, one reinterprets a certain field $A_{\mu}$ within the domain of other fields $A_{\mu I}, A_{\mu}\to \{A_{\mu I}\},\text{where } I \in 1...N$. This extension makes physics manifest the part in the whole. Develop new physical properties.  
 
\subsection{Fields set basis} 
${}$\indent

The possibility of expanding the gauge parameter for a systemic relationship, results that, the corresponding physical properties can be seen under different fields basis. Similarly, as physical laws are independent of reference systems in space-time, AGT according to Salam-Kameguchi theorem can be analyzed under different reference of fields such as $\{A_{\mu I}\}, \{D_{\mu}, X_{\mu}^{i}\}, \{G_{\mu I}\}$ [6]. 

Thus, the basis $\{A_{\mu I}\}$ associate a plurality of fields transforming under the same parameter, $\{A'_{\mu I}\} = \{A_{\mu I}\} + k_{I}\partial_{\mu}\alpha$. The basis $\{D_{\mu},X_{\mu}^{i} \}$ is useful for analizing explicitly the gauge invariance. Nevertheless, it is at fields basis $\{G_{\mu I}\}$, where physics is done. The physical fields are those that diagonalize the equations of motion. Defined as the poles at two point Green's function.

A rotation from the constructor basis has to be implemented to define the physical fields $\{G_{\mu I}\}$. In terms of these fields that the corresponding measurable entities of the model must be defined, as the corresponding electric and magnetic fields. At Appendice A, one studies the so-called constructor basis $\{D_{\mu}, X_{\mu }^{i}\}$. It is the simplest way to reveal on the physical abelian set gauge invariance.

\subsection{Gauge transformation}
${}$\indent

The fields set approach is being studied. It provides quanta under the set action. The physical basis $\{G_{\mu I}\}$ is defined in terms of the constructor basis$\{D_{\mu}, X_{\mu i}\}$ through the $\Omega$ matrix rotation [7]. From Appendix A, one writes

\begin{equation}
D_\mu = \Omega_{1I}G_\mu^I 
\end{equation}

\begin{equation}
X_{\mu i} = \Omega_{iI}G_\mu^I, \text{ i= 2,...,N}
\end{equation}
where $\Omega$ is a transformation matrix defined in terms of lagrangian coefficients. It does not depend on spacetime. It corresponds to the transverse sector diagonalization. Index i means the flavour  related to $i=2,...,N$.

Thus, the physical basis is a $\Omega$ transformation from constructor basis. The matrix $\Omega$ has the property of being invertible
\begin{equation}\label{omega property}
    {\Omega_{IK}\Omega_{KJ}^{-1} = \delta_{IJ}}
\end{equation}

It gives,
\begin{equation}
\begin{pmatrix}
G_{\mu 1}\\
G_{\mu 2}\\
\vdots\\
G_{\mu N} 
\end{pmatrix}=\begin{pmatrix}
\Omega^{-1}_{11} & \Omega^{-1}_{12} &\dots & \Omega^{-1}_{1N}\\
\Omega^{-1}_{21} & \Omega^{-1}_{22} &\dots & \Omega^{-1}_{2N}\\
\vdots&\vdots&\ddots&\vdots\\
\Omega^{-1}_{N1} & \Omega^{-1}_{N2} &\dots & \Omega^{-1}_{NN}\\
\end{pmatrix}\begin{pmatrix}
D_\mu\\
X_{\mu 2}\\
\vdots\\
X_{\mu N}
\end{pmatrix}
\end{equation}
resulting

\begin{eqnarray}
    {G_{\mu I} = \Omega^{-1}_{I1}D_\mu + \Omega^{-1}_{Ii}X_{\mu}^i}
\end{eqnarray}

Thus the corresponding gauge transformation for every physical field envolved in the abelian set is

\begin{equation}\label{G'muI}
    G_{\mu I}' = G_{\mu I} + \Omega^{-1}_{I1}\partial_{\mu}\alpha
\end{equation}
where eq. (\ref{G'muI}) is expressing that each field inserted in the fields set $\{G_{\mu I}\}$ is individualized by its own gauge transformation and with their transversal poles diagonalized. It is identified as the symmetry of difference.

\section{Gauge invariance}
\subsection{Fields strengths}
${}$\indent

The interwined fields $\{G_{\mu I}\}$ determine various fields strengths from eq. (\ref{G'muI}). They will have granular and collective nature, with antisymmetric and symmetric properties. Considering first, the antisymmetry sector, one rewrites Apendice A as below.

\vspace{0.5cm}
{\large\bf A. Antisymmetric sector}

The corresponding granular fields strengths are given by 

\begin{equation}\label{GImunu}
{G^I_{\mu\nu}} = \partial_{\mu}{G^I_{\nu}}- \partial_\nu {G^I_{\mu}}\nonumber
\end{equation}
whose gauge invariance is supported by eq. (\ref{G'muI})
\begin{equation}
    G^{I'}_{\mu \nu} = G^{I}_{\mu \nu}
\end{equation}

Studying the collective field strength, one gets from Apendice A

\begin{equation}
    z_{[\mu \nu]} \equiv \gamma_{[IJ]}G_{\mu}^IG_{\nu}^J, \textbf{ }\gamma_{[IJ]} \equiv \gamma_{[ij]} \Omega^{i}_I\Omega^{j}_J
\end{equation}
The collective antisymmetric field gauge invariance

\begin{equation}
z_{[\mu\nu]}' = \gamma_{[IJ]} {G^I_\mu}'{G^J_\nu}' = z_{[\mu \nu]}\nonumber
\end{equation}
is derived on eq. (\ref{omega property}) relationships
\begin{eqnarray}
    &&\Omega^{i}_{I}\Omega^{j}_{J}\Omega^{-1J}_{1} =\Omega^{i}_{I}\delta_{j1}=0, \text{ where } j=2...N,\nonumber
    \\
    &&\Omega^{i}_{I}\Omega^{j}_{J}\Omega^{-1I}_{1} =\Omega^{i}_{I}\delta_{i1}=0, \text{ where } i=2...N,\nonumber
    \\
    &&\Omega^{i}_{I}\Omega^{j}_{J}\Omega^{-1I}_{1}\Omega^{-1J}_{1} = \delta_{i1}\delta_{j1} = 0.
\end{eqnarray}
The collective fields strength is generalized by introducing other potential fields associations [8],
\begin{eqnarray}
    s_{\mu\nu} = s_{IJ}G^{I}_{\mu}G^{J}_{\nu}, \quad s_{IJ}=\sum^{N}_{i,j=2}\Omega_{iI}\Omega_{jJ},
\end{eqnarray}
\begin{eqnarray}
    t_{\mu\nu} = t_{IJ}G^{I}_{\mu}G^{J}_{\nu}, \quad t_{IJ}=\sum^{N}_{K,L=1}\gamma_{KI}\gamma_{LJ},
\end{eqnarray}
\begin{eqnarray}
    u_{\mu\nu} = u_{IJ}G^{I}_{\mu}G^{J}_{\nu}, \quad u_{IJ} = \sum_{i,K}\Omega_{iI}\gamma_{KJ}
\end{eqnarray}
which produces the generic gauge invariant collective field,
\begin{eqnarray}
e_{\mu\nu} = z_{\mu\nu} + t_{\mu\nu} + s_{\mu\nu} + u_{\mu\nu}
\end{eqnarray}
where
\begin{eqnarray}
e_{\mu\nu} = e_{IJ} G^{I}_{\mu} G^{J}_{\nu}
\end{eqnarray}
with
\begin{eqnarray}\label{eIJ}
e_{IJ} = \gamma_{IJ} + s_{IJ} + t_{IJ} + u_{IJ}
\end{eqnarray}
where
\begin{eqnarray}
&&e_{IJ} = \sum^{N}_{K,L = 1} \sum^{N}_{\overset{i,j}{k,l} =2} \left(1 + \gamma_{ij} + \gamma_{kj}\Omega_{kK}\right) \Omega_{iI}\Omega_{jJ}+\nonumber\\
&&+\gamma_{ki}\gamma_{lj}\Omega_{kK}\Omega_{lL}\Omega_{iI}\Omega_{jJ}.
\end{eqnarray}

Thus, the abelian fields set constructs the generic antisymmetric gauge invariant tensor

\begin{eqnarray}\label{2.21}
Z_{[\mu\nu]} =a_IG^I_{\mu\nu}+e_{[\mu\nu]} 
\end{eqnarray} 
where
\begin{eqnarray}
a_I=d\Omega_{1I} + \alpha_i \Omega_{iI}
\end{eqnarray} 
and

\begin{equation}
    Z^{'}_{[\mu \nu]} = Z_{[\mu \nu]}.
\end{equation}

{\large\bf B. Symmetric sector}

The constructor basis produces the generic gauge invariant field strength
\begin{equation}\label{Generic Tensor Simetric}
    Z_{(\mu\nu)} = \beta_i S^i_{\mu\nu}+\rho_i g_{\mu\nu}S^{\alpha i}_\alpha+\gamma_{(ij)}X^i_\mu X^j_\nu+ \tau_{(ij)} g_{\mu\nu} X^i_\alpha X^{\alpha j}
\end{equation}
where
\begin{eqnarray}
    S^i_{\mu\nu} =\p_\mu X^{i}_\nu+\p_\nu X^{i}_\mu =\Omega^i_I S^I_{\mu\nu}
\end{eqnarray}
and
\begin{equation}
    S^{\alpha i}_{\alpha} = \Omega^i_I S_\alpha^{\alpha I}
\end{equation}

The corresponding collective symmetric field strength is
\begin{eqnarray}
     z_{(\mu \nu)} \equiv{\gamma_{(ij)}X_\mu^i X_\nu^j} &=& \gamma_{(IJ)} G_{\mu I} G_{\nu J}
\end{eqnarray}
Similarly,
\begin{equation}
    \omega^\alpha_\alpha \equiv \tau_{(ij)}X^i_\alpha X^{\alpha j} = \tau_{(IJ)}G_{\alpha}^{I}G^{\alpha J}
\end{equation}

Generalizing the symmetric collective field strenght,
\begin{eqnarray}
e_{(\mu\nu)} = z_{(\mu\nu)} + s_{(\mu\nu)} + t_{(\mu\nu)} + u_{(\mu\nu)}
\end{eqnarray}
where
\begin{eqnarray}
e_{(\mu\nu)} = e_{(IJ)} G_{\mu I} G_{\nu J}
\end{eqnarray}

Substituting at eq. (\ref{Generic Tensor Simetric}), in terms of physical fields, one gets the following gauge invariant symmetric tensor

\begin{equation}
    Z_{(\mu\nu)} = {\beta_I} S^I_{\mu\nu}+{\rho_I} g_{\mu\nu} S_\alpha^{\alpha I}+\mathbf{e}_{(\mu\nu)}+ g_{\mu\nu}\omega^\alpha_\alpha 
\end{equation}
where
\begin{eqnarray}
     S^I_{\mu\nu} &=& \p_\mu G^I_\nu+\p_\nu G^I_\mu\\
     \mathbf{e}_{(\mu\nu)} &=& \mathbf{e}_{(IJ)}
     G^I_\mu G^J_\nu\\
     \omega_{(\mu\nu)} &=& \tau_{(IJ)} G^I_\mu G^J_\nu\\
\end{eqnarray}
with the coefficients
\begin{eqnarray}
    &&\beta_I = \beta_i\Omega^i_{I} \text{  ;  } \rho_I = \rho_i\Omega^i_{I}
    \\
    &&\mathbf{e}_{(IJ)} = \gamma_{(IJ)} + s_{(IJ)} + t_{(IJ)} + u_{(IJ)} 
    \\
    &&\tau_{(IJ)} = \tau_{(ij)}\Omega^i_{I}\Omega^j_{J}.
\end{eqnarray}
\subsection{Constructivist Abelian Lagrangian}
${}$\indent
A whole quanta process is expected from fields strengths through a constructivist lagrangian. There is an abelian grouping fields to be explored. A physics assembled by whole gauge symmetry. Introduce on fields agglomerations and stipulate their corresponding quanta. Derive properties inserted in a whole.

The abelian constructive lagrangian will be studied in terms of fields strenghts. It yields the following expression

\begin{equation}
    \lag = Z_{[\mu \nu]}Z^{[\mu \nu]}+Z_{(\mu \nu)}Z^{(\mu \nu)}+\eta Z_{[\mu \nu]}\Tilde{Z}^{[\mu \nu]}+m^2_{II} G^I_\mu G^{\mu I} + \xi_{IJ}(\partial_{\mu}G^{\mu I})(\partial_{\nu}G^{\nu J})\label{equation 3.29}
\end{equation}
 including the dual tensor $\Tilde{Z}^{[\mu \nu]} = \epsilon_{\mu \nu \rho \sigma}Z^{[\rho \sigma]}$, with the term $Z_{[\mu \nu]}\Tilde{Z}^{[\mu \nu]}$, which is not exactly a total derivative. It gives, 

\begin{eqnarray}
    Z_{[\mu \nu]}\Tilde{Z}^{[\mu \nu]} = \epsilon^{\mu \nu \rho \sigma}\left\{2a_{I}\left[\partial_{\mu}(G^{I}_{\nu}Z_{[\rho \sigma]}) -G_{\nu}^{I}\partial_{\mu}z_{[\rho \sigma]}\right] + z_{[\mu \nu]}Z_{[\rho \sigma]}\right\}
\end{eqnarray}
Eq. (3.30) is called as semitopological lagrangian [9].

Splitting the fields strengths, one gets the following non-linear abelian Lagrangian, in terms of physical fields, 

\begin{eqnarray}
    &&\lag (G)= a_{I}a_{J}G_{\mu \nu I}G^{\mu \nu J} + a_{I}G_{\mu \nu I}\mathbf{e}^{[\mu \nu]} + \mathbf{e}_{[\mu \nu]}\mathbf{e}^{[\mu \nu]}\nonumber
    \\
    &&+\beta_{I}\beta_{J}S_{\mu \nu I}S^{\mu \nu J} + 2\beta_{I}\rho_{I}S_{\alpha I}^{\alpha}S^{\beta I}_{\beta} +2\beta_{I}S_{\mu \nu}^{I}\mathbf{e}^{\mu \nu}\nonumber
    \\
    &&+2\beta_{I}S^{\alpha I}_{\alpha}\omega_{\beta}^{\beta} + 17\rho_{I}\rho_{J}S^{\alpha I}_{\alpha}S^{\beta J}_{\beta} +2\rho_{I}S^{\alpha I}_{\alpha}\mathbf{e}^{\alpha}_{\alpha}\nonumber
    \\
    &&22\rho_{I}S^{\alpha I}_{\alpha}\mathbf{\omega}^{\alpha}_{\alpha}+16\omega_{\alpha}^{\alpha}\omega_{\beta}^{\beta}  +  \partial_{\mu}\{2\epsilon^{\mu \nu \rho \sigma}a_{I}G_{\nu}^{I}Z_{[\rho \sigma]}\}+\nonumber
    \\
    &&a_{I}\mathbf{e}_{[KL]}\epsilon^{\mu \nu \rho \sigma}G_{\mu}^{I}G_{\rho}^{K}\partial_{\nu}G_{\sigma}^{L}+2a_{K}\mathbf{e}_{[IJ]}\epsilon^{\mu \nu \rho \sigma}G_{\mu}^{I}G_{\nu}^{J}\partial_{\rho}G_{\sigma}^{K}\nonumber
    \\
    &&\mathbf{e}_{[IJ]}\mathbf{e}_{[KL]}\epsilon^{\mu \nu \rho \sigma}G_{\mu}^{I}G_{\nu}^{J}G_{\rho}^{K}G_{\sigma}^{L}
\end{eqnarray}

Separating in pieces

\begin{eqnarray}\label{Lagrangian fundamental}
    \lag (G)= \lag_{K} + \lag_{I} + \lag_{st} + \lag_{M} + \lag_{GF}\label{equation 3.32}
\end{eqnarray}

 The kinetic term is given by
\begin{equation}
    \lag_{K} = a_{I}a_{J}G_{\mu \nu I}G^{\mu \nu J} +\beta_{I}\beta_{J}S_{\mu \nu I}S^{\mu \nu J}+2\beta_{I}\rho_{I}S_{\alpha I}^{\alpha}S^{\beta I}_{\beta}
\end{equation}

The interaction sector contains the abelian trilinear terms,
\begin{equation}
    \lag^3_I = a_{IJK}\p^\mu G^{\nu I}G^{J}_\mu G^{\nu K} + 4b_{IJK}\p_\mu G^{I \mu}G^J_\nu G^{k\nu}\label{equation 3.34}
\end{equation}
or 
\begin{equation}
    \lag^3_{I} = (a_{IJK} + b_{JIK}+b_{KJI})\partial^{\mu} G^{\nu I}G_{\mu}^{J}G_{\nu}^{K}
\end{equation}
and abelian four linear terms,
\begin{equation}\label{Lagrangian quad}
    \lag^4_I = a_{IJKL}  G_{\mu}^{I}G_{\nu}^{J}G^{\mu K}G^{\nu L} + b_{IJKL} G_{\mu}^{I}G^{J \mu}G^{ K}_\nu G^{L\nu}
\end{equation}

The semitopological term also receives tri and quadrilinear contribuitions,

\begin{eqnarray}
    \lag^{3}_{st} = a_{I}\mathbf{e}_{[KL]}\epsilon^{\mu \nu \rho \sigma}G_{\mu}^{I}G^{K}_{\rho}\partial_{\nu}G_{\sigma}^{L} + 2a_{K}\mathbf{e}_{[IJ]}\epsilon^{\mu \nu \rho \sigma}G_{\mu}^{I}G^{J}_{\nu}\partial_{\rho}G_{\sigma}^{L}
\end{eqnarray}
and
\begin{eqnarray}
    \lag^{4}_{st} = \mathbf{e}_{[IJ]}\mathbf{e}_{[KL]}\epsilon^{\mu \nu \rho \sigma}G_{\mu}^{I}G_{\nu}^{J}G_{\rho}^{K}G_{\sigma}^{L}\label{equation 3.38}
\end{eqnarray}

In order to get a better underdstanding on the Lagrangian spectroscopy, one should make a coefficients decomposition through Young Tableaux. It gives

\begin{equation*}
    a_{IJ} = a_{[IJ]} + a_{(IJ)}
\end{equation*}
where
\begin{eqnarray}
    a_{(IJ)} = \Tilde{a}_{IJ} + \frac{1}{N}\delta_{IJ}tr(a)
    \textbf{, }tr(\Tilde{a}_{IJ}) =0,
\end{eqnarray}
and

\begin{eqnarray}\label{aIJK}
    &&a_{IJK} = a_{[IJK]}+a_{[I(J)K]}+a_{[I(J)K]}+a_{[I(J)K]}+a_{(IJK)}
\end{eqnarray}
where

\begin{eqnarray}
    a_{(IJK)} = \Tilde{a}_{(IJK)} + \frac{1}{N+2}(\delta_{IJ}a_{K} + \delta_{JK}a_{I} + \delta_{IK}a_{J}),
\end{eqnarray}
and
\begin{eqnarray}
    a_{IJKL} &= a_{(IJKL)} + a_{(IJ[K)L]} + a_{(I[J)(K]L)} + a_{[IJ(K]L)} + a_{[IJKL]},
\end{eqnarray}
as
\begin{eqnarray}\label{aIJK}
    &&a_{IJKL} = a_{[IJKL]}+a_{[I(J)KL]}+a_{[I(J)KL]}+a_{[I(J)KL]}+a_{[IJKL]}
\end{eqnarray}
where
\begin{eqnarray}
    a_{(IJKL)} = \Tilde{a}_{(IJKL)} + (\delta_{IJ}a_{KL} + \delta_{KL}a_{IJ} + \delta_{IK}a_{JL} + \delta_{JL}a_{IK}),
\end{eqnarray}
where

\begin{eqnarray}
    a_{(IJ[K)L]} = \Tilde{a}_{(IJ[K)L]} + \delta_{IJ}a_{KL},
\end{eqnarray}

\begin{eqnarray}
    a_{(I[J)(K]L)} =  \Tilde{a}_{(I[J)(K]L)} + \delta_{IL}a_{JK},
\end{eqnarray}

\begin{eqnarray}
    a_{[IJ(K]L)} = \Tilde{a}_{[IJ(K]L)} + \delta_{IL}a_{JK}
\end{eqnarray}
Notice that eq.(3.33-3.36) shows a soft nonlinearity. Given than the corresponding $\Tilde{a}_{IJK}, \Tilde{b}_{IJK} ...$ can be taken zero separately without breaking gauge simmetry, it is possible, for instance, to reduce the interaction term for $\Tilde{a}_{IJK}$.

The fields set composes masses without requiring Higgs mechanism [10]. A mass term more near to energy and Anderson prescription [11] than considering degress of freedon transference as at symmetry breaking. It gives,

\begin{eqnarray}\label{completely lagrangian}
    \lag_m &=& m^2_{ij} X^i_\mu X^{\mu j}\no
    = {m^2_{II}}G^I_\mu G^I_\nu\no
\end{eqnarray}
or
\begin{equation}
        \lag_{m}(G) = m^2_{I I}G^I_{\mu} G^{\mu I}
\end{equation}

Given the existence of just one gauge parameter, there is just one gauge fixing. It is written as a linear combination of the fields $G_{\mu I}$. From Appendix A, it yields

\begin{equation}
    \lag_{GF} = \xi_{IJ}\partial_{\alpha}G^{\alpha I}\cdot\partial_{\beta}G^{\beta J}
\end{equation}

Thus, the abelian constructivist lagrangian is constituted. Eq. (\ref{Lagrangian fundamental}) contains independent gauge invariants pieces. Their physical relevance are terms providing own physics. Eqs. (3.33-3.38)  develop isolated physical conglomerates. Each of them works as a source to do nonvirtual physics.
\newpage

\section{Physical realities and potentialities}

${}$\indent
Reality in gauge theory is given by gauge scalars. Diverse from Yang-Mills case [12], the abelian constructivist lagrangian is separated in different sectors where each one is gauge invariant, as eq. (3.32) shows. It generates diverse subunits of reality. They contain their own physics with potentialities to be studied.  

The constructivist abelian lagrangian is not monolithic. Usually gauge theories develop lagrangians which are gauge invariant as a whole, i.e, their separated pieces are virtual. As a new aspect, the fields set physics is constituted by blocks where each of them englobes own opportunities. Each one is individually able to do real physics. We list below these independent structures. Gauge scalars subunits as source for interdependent electromagnetic energy. They are

\begin{align}
&G_{\mu \nu I}G^{\mu \nu I}, &  &\epsilon_{\mu \nu \lambda \sigma}G^{\mu\nu I}G^{\lambda \rho I},  &  &a_{IJ}S_{\mu \nu}^{I}S^{\mu \nu J}, &  &a_{IJ}\epsilon_{\mu \nu \lambda \rho}S^{\mu \nu I}S^{\lambda \rho J},\nonumber
\\
&a_{J}G_{\mu \nu}^{I}S^{\mu \nu J} &  &a_{J}\epsilon_{\mu \nu \lambda \rho}G^{\mu \nu I}S^{\lambda \rho J}   &  &G_{\mu \nu}^{I}\mathbf{e}^{[\mu \nu]}  &      &\epsilon_{\mu \nu \lambda \rho}F^{\mu \nu I}\mathbf{e}^{[\lambda \rho]},\nonumber
\\
&\mathbf{e}_{[\mu \nu]}\mathbf{e}^{[\mu \nu]},   &  &\epsilon_{\mu \nu \lambda \rho}\mathbf{e}^{[\mu \nu]}\mathbf{e}^{[\lambda \rho]},     &  &a_{I}S_{\mu \nu}^{I}\mathbf{e}^{(\mu \nu)},   &  &a_{I}\epsilon_{\mu \nu \lambda \rho}S^{¨\mu \nu I}\mathbf{e}^{(\lambda \rho)},\nonumber
\\
&\mathbf{e}_{(\mu \nu)}\mathbf{e}^{(\mu \nu)},  & &a_{I}\epsilon_{\mu \nu \lambda \rho}\mathbf{e}^{(\mu \nu)}\mathbf{e}^{(\lambda \rho)},
\end{align}

Nevertheless in terms of quanta and interactions it is more appropiate to write eq. (3.29) in terms of eqs. (3.32). They will split on fields associations potentialities. Take as example a three linear piece as $a_{IJK}\partial^{\mu}G^{\nu I}G_{\mu}^{J}G^{K}_{\nu}$. It will produce its own real physics. Their associations will contribute for a real trilinear nonlinear term and for a nonvirtual vertice set of  Feynman graphs.

The constructivist lagrangian is rewritten at table below. Diverse scalars terms are catalogued. Each one contains parameters in front of it. They are called as abelian symmetry free coefficients. Originated at Appendix A and transcripted to eq. ( \ref{equation 3.29}) they can take any value without violating the gauge invariance. The number of free coefficients associated to each gauge invariant term constitute the so-called volume of circumstances.

\newpage
\begin{table}[htbp]
  \centering
   \caption{Relationship between scalar gauge and Nº of free coefficients}
    \begin{tabular}{l|l|l}
    \hline
    Scalar &  Free Coefficients & Nº of Free Coefficients\\
    \hline
    $G_\mu^I \Box \theta^{\mu\nu} G_\nu^J$ & $d^2\Omega_{1I}\Omega_{1J}+2d\alpha_i\Omega_{1I}\Omega^i_{J}+\Omega_{J}^j$ & $\frac{4N(N+1)+N^2(N-1)}{8}\cdot$ \\ &$+ 2(\alpha_i\alpha_j\beta_i\beta_j)\Omega_{I}^i$ &$\cdot[N^4+5N^2+2(N^2-(N-1))]$ \\
    \hline
    $G_\mu^I \Box \omega_{\mu\nu} G^{L \nu}$ & $(\beta_i \beta_j+2\rho_i\beta_j+\rho_i\rho_j)\Omega_{I}^i\Omega_{J}^j$ &$ \frac{N^3(N^2-1)}{16}[N^2(N+1)+2]$  \\
    \hline
      $G_\mu^I G^{I\mu}$ & $m^2_{ij}\Omega_{I}^i\Omega_{I}^j$ &$\frac{N^3(N-1)^3}{2}$\\
    \hline
    $\p^\mu G^{\nu I} G^{J}_\mu G^{\nu K}$ & $d\mathbf{e}_{[ij]}\Omega_{1i} \Omega^i_J \Omega^j_K+ $ & $N^3(N-1)^3\cdot$\\ &$+(\alpha_i\mathbf{e}_{[kj]}+\beta_i \mathbf{e}_{[kj]})\Omega_{I}^i\Omega_{J}^j\Omega_{K}^k$ & $\cdot[(N-2)+N(N-1)]$\\
    \hline
    $\p_\mu G^{\mu I} G^{J}_\nu G^{K \nu}$ & $(\beta_i\tau_{(kj)}+\rho_i\gamma_{(kj)}+\rho_i\tau_{(kj)})\Omega_{I}^i\Omega_{J}^j\Omega_{K}^k$ & $\frac{3N^9(N-1)^5}{2}$\\
    \hline
    $G_\mu^I G^{J}_\nu G^{\mu K}  G^{\nu L}$ & $(\mathbf{e}_{(ij)}\mathbf{e}_{(kl)}+\mathbf{e}_{(ij)}\mathbf{e}_{(kl)})\Omega_{I}^i\Omega_{J}^j\Omega_{K}^k\Omega_{L}^l$ &$\frac{N^4(N-1)^4}{2}[N(N-1)+3]$ \\
    \hline
    $G_\mu^I G^{J\mu} G^{K}_\nu  G^{L \nu}$ & $(\mathbf{e}_{(ij)}\mathbf{e}_{(kl)}+\tau_{(ij)}\tau_{(kl)})\Omega_{I}^i\Omega_{J}^j\Omega_{K}^k\Omega_{L}^l$ &  $\frac{N^4(N-1)^4}{8}[3N(N-1)+2]$\\
    \hline
    \end{tabular}%
  \label{tab: Gmn}
\end{table} 

Table (4.1) expresses the reality of the abelian constructivist lagrangian subunits. It catalogues each physical term potentiality. Eq. (3.29) is not a rigid physics. Their fields associations constitute subunits of reality, independent terms due to their gauge invariances and interdependent by sharing common fields. Their potentiality means the volume of circumstances propitiated by gauge symmetry to each gauge scalar term. So, given by the number of free coefficients associated to every scalar term, a physics with free will is introduced. A new way to express energy.

Summarizing, the constructivist lagrangian contain terms with own realities and potentialities. And so, the phenomena may be analyzed by different lagrangian pieces where each one contains its own reality and potentiality. Physical entities as energy, mass, coupling constants, conservation laws, Feynman graphs will be depending on each of these separated lagrangian blocks.

\section{Associative physics}
${}$\indent
Constructivist lagrangian establishes physics under fields set. The grouping physics. A significance under associativity. The quantum is no longer an enterprise from itself, but of a set. A perspective where the corresponding properties will be ruled under environment.

Physics laws derived from fields environment is a study to be considered. The fields set $\{G_{\mu I}\}$ does not consider ultimate constituents as definitive. They produce an environmental physics. Consequently, the search for elementarity is not only through reductionist subunits, but also, by inserting parts in the whole.  Instead of studying an atomist gauge theory like the Yang-Mills [13], whose objective is to reveal the fundamental interactions, it turns out to consider an integral physics A constructive lagrangian where particles are described by a third quantum depending on fields set. 

Nature is more than playing with ultimate constituents. It also processes the associativity through conglomerate's determination. The whole-part duality. A complementarity relating the quantum and to the whole. Nature also works as a collective manifestation. A  quantum system physics to be studied beyond four interactions.

\subsection{Ultimate constituents and two quantum formations}
${}$\indent

Subunit and environment are two nature constructions. The atomist and grouping behaviours. Physics followed the first, while biology found the second. The physics postulate has been that subunit precedes wholeness. It concerns on how phenomena are atomistically constituted. Consider on Big Bang and the formation of particles. Let aside the existence of conglomerates.

Nevertheless nature provides groupings. A set physics supported by general principles as associativity, confinement, complexity. Introducing a relationship between the part and the whole. For instance, particles getting together at eightfoldway and atoms at periodic table. They are relating how systems are constituted from constituents. Quarks and protons creating groupings. Showing that, subunits are not only working for composing particles, as gathering them into conglomerates. 

Thus, there are two ways for expressing the nature of things. Possibilites on quantum formation. Reductionistic and antireductionistic. Procedures from ultimate constituents. On one hand subunit, another set. The first one, based on building blocks (as quarks and leptons); other, on associations  (conglomerates). Two performances for expressing the quantum.

The 1940s and 50s decades led physics to witness the proliferation of elementary particles. At that time, it was considered as a crisis in physics such diversity of particles. However, the discovery of the omega minus particle in 1964, definitively, came to show that these particles were not isolated, but associated. There was an eightfoldway grouping physics promoted by ultimate constituents called quarks.

Thus, at 1960s physics encountered a relationship between the greek atom and the notion of set. After Planck and quantum mechanics, came out the existence of a new type of quantum. The eightfoldway showed that, proton, neutron and so forth are not isolated particles. They are formed by quarks, and also, associated by quarks. The Quark Model provides particles and patterns of particles. Showing that, nature contains ultimate subunits as quarks, and, develops conglomerates unities as eightfoldway representations. Consequently, instead of isolated particles, nature produces particles in associations.

A third quantum physics emerges. Particles physics is not a zoo of dissociated particles, but, under groupings associations. Eightfoldway shows that the ultimate constituents produce particles and environment of particles. Particles and systems are created. Therefore, beyond the greek atm there another perspective to be token. Ultimate constituents are formatting the environmental physics.

There is an unexplored physics. The relationship between ultimate constituent and conglomerate. A new elementarity has to be proposed. The quantum related to the system which is involved. Composed by ultimate constituents and respective grouping physics this quantum is more than a property of itself. The part in the whole turns up as the third quantum configuration. In reference to biology, this whole quantum will be denominated as variety. The elementarity constitued by the fields set environment.

Nevertheless, particle physics has always been impregnated by atomism. The Quark Model subproducts, protons and neutrons, appeared constituted by quarks but as dissociated particles. Physics just study their strong interactions. Something is lost. Incorporate their associative physics. Consider the quantum dependence on ultimate constituents and whole associations. 

The Quark Model atomist philosophy did not registered the conglomerated physics. The search for the greek atom over the last 2500 years has diverted physics from set notion. As result, current physics interprets nature under interactions and not associations. Subtracting the meaning that, the universe is an association derived from ultimate constituents, and, as a natural order, providing the development of patterns.

Physics has to include the existence of variety. A new elementarity under the set action. The quantum associated to a fields environment. A system with an own physics to be expressed generating distinct energy packets. The proton identified as an eightfoldway variety. There are new physical properties to be investigated under the conglomerate view.

Physisc has to express the variety. Based on associativity, eq. (\ref{equation 3.32}) definies a new energy packet. An antireductionist quantum is constituted. A definition where physics contains another perspective for describing the phenomena. The relationship between ultimate constituents and conglomerates cause another type of procedure. Near to biology, it will identified as constructivist physics.

\subsection{Constructivist physics}
${}$\indent

A constructivist physics is identified. A physics beyond atomism. Since 1900 with Planck granularity and 1927 with quantum field theory, the quantum world became the lecture of greek atm. However, the atomism is not the last quantum performance. The principles of associativity, confinement and complexity are requiring the existence of a grouping physics. Physics under the set action. Consequently, the part in the whole generates environmental properties to be explored. 

A third quantum should exist. A comprehension about the quantum behaviour within of the whole proceed. Consider set physics constituted by Lorentz group and gauge symmetry. While the usual reductionist methodology defines particles interacting with each other mediated by gauge bosons, the antireductionist way will relate them through a systemic behaviour. Consider each Lorentz group representation associated with a fields set. For example, in the vector case, the $A_{\mu}$ field representation $(\frac{1}{2}, \frac{1}{2})$ is extended to the fields set $\{A_{\mu I}\}$ [14]. 

Gauge symmetry is also reinterpreted. Instead of establishing just fundamental interactions, its new physical proposal is to assemble fields. Establish nature in associations through antireductionist gauge symmetry. An environmental physics emerges. Quanta insert in the whole is what the antireductionist gauge symmetry provides. Given eq. (\ref{equation 3.32}), the fields set $\{G_{\mu I}\}$ provides $(2N-1)$ interconnected quanta to be compiled [15]. 

Thus, the constructivist physics interprets nature as a whole atom. Establish a systemic physics by groupings the Lorentz transformations $A_{\mu I}=\Lambda_{\mu}^{\nu}A_{\mu I}$ and the gauge symmetry by eq. (2.6). A whole behaviour to be analyzed. Instead of looking for isolate subunits address the principle of wholeness. Elementary particles and respective quantum numbers constituted from the whole which are envolved. Whole-part duality and quanta network constitute a new physics.

The content of grouping physics provokes a new understanding. A performance where the quantum is not expressed isolatedely. A new construction from ultimate constituents is identified. Constructivist lagrangian provides a systemic quantum packet. The whole quantum associativity. Ultimate constituents forming the whole-part duality are causing an environmental physics.

\subsection{Quantum system: environment, variety and functionality}
${}$\indent 

A perspective not well exploited by quantum field theory is that physics includes the grouping manifestation. An integral activity moved not only by fundamental interactions, but firstly, under associativity. Dirac equation relating electrons and positrons and eightfoldway are exemplifying on this associative behaviour. They make physics join to biology on nature constructivism. Specie in biology and whole quantum in physics share the meaning of environment. Based on that, both should be subjects of nature associative behaviour.

Biology managed to privilege environment, variety and functionality over ultimate constituents and interactions. Biology takes variety from environment. Darwin, through the differentiation of species, found the evolutionary timeline. Introduced the notion of adaptive variety. However, given that biology and physics are natural sciences, this line of thought should be extended into physics. As biology, physics working out as a subject proposing environment under fields set. Consider that physical bodies, as particles, contain a behaviour major than just interactions.

The interconnection between physics and biology should be analysed. Darwin in 1859 understood the relationship between species (varieties) and environment [16]. Similarly, Gell-Mann and Ne'eman in 1961 joined particles in eightfoldway groupings [17]. Later in 1937, Dobzhansky introduced the relationship between genes, environment and varieties [18]. Similarly, in 1964 Gell-Mann and Zweig introduced quarks at group representations 3, 3* and developing conglomerates through representations 8,10,15,... [19]. Demonstrating that, physics also contains a relationship between ultimate constituents and grouping. Formatting particles as varieties. Therefore, another consequence from quarks is working as subunits constituting varieties.  

Thus, a third phase on quantum story is the variety physics. After granular and wave-particle behaviours the performance is on the quantum association. The Quark Model generating interlinked composite particles through conglomerates brings a new perfomance. Ultimate constituents forming a set action and introducing as whole quantum. It shows unlike restricting physics to isolated reductionist processes, one gets that, environment, variety and functionalities are establishing a new possibility to do physics. A third type of energy packet processed with antireductionist properties.

A new situation is stated. The association physics enlarges phenomena beyond the interactive atomism. The relationship between ultimate constituent and environmental varieties generates a next perspective. Eq. (2.6) symmetry introduces a fields set and eq. (3.23) a set action. Interaction is enlarged  to interrelationship. Grouping arrangements producing antireductionist properties beyond isolated particles. The atomist four interactions are prolonged to antireductionist quantum functionalities.

\subsection{Grouping properties}
${}$\indent

A new physics appears. The atomism is extended to constructivism through gauge symmetry. A third quantum is established under a whole field theory. Inserted in the whole adds new physicalities. The fields set produces relationships beyond the atomist perspective. Maintaining the original reductionism symmetry, and so, preserving the four fundamental forces a quantum system is constituted. New physical relationships are performed. Antireductionist properties depending on conglomerate are produced.

The physics of a single particle is extended for a system. A field $G_{\mu I}$ is incorporated at $\{G_{\mu I}\}$. A quantum system is established. An extension where the single particle physics is preserved. The covariant derivative is preserved, and so, the original atomist interaction is not modified. But, new relationships based on interlinked fields are developed. The aggregation enlarges physical laws on interaction, induction and connection. That is the theme to be explored.

Three types of physics emerge from the set action. Associativity, set determinism, complexity. Through associativity appears set, diversity, interdependence, nonlinearity, chance. The resulting conglomerate unity introduces a set transformation physics. A set determinism under directive and circumstance, growth, evolution, emergence. Features which will constitute a multitude acting together. Promoting complexity.

Associativity introduces the grouping behaviour. The antireductionist gauge symmetry provides the whole quantum system under difference and regularity. Whole-quantum duality and quanta network are constituted. Varieties under diversity. Kinematically, an interconnectivity identified by quantum numbers. The respective quanta are not independent of each other, but interrelated by common quantum numbers. Dynamically, potential fields interdependences are expressed through equations of motion, Faraday laws, conservation laws, Noether laws, constructive Lorentz force.

Thus, before considering the interaction between particles, we must observe their interrelationships. For example, when assuming the photon and $Z_{0}$ inserted in the same set, it turns out that, before supposing them as independent carriers of the electromagnetic and weak interactions, we must start from the fact that they are associated particles. Making electromagnetism and weak interaction interrelated.

Nonlinearity and chance are also properties beyond the quantum atomism. The first one, establishes a self organizing system. A dynamics without depending on external sources. The concept of chance is derived from the potentialities that each gauge scalar term produce. It yields the physics of opportunities. The quantum is no more a fixed structure, but an entity which properties are under chance. Decisions depending on each lagrangian gauge invariant term. Even at abelian case, tri-and-quadrilinear chances are expressed at eqs.(\ref{equation 3.34}-\ref{equation 3.38}). Nonvirtual opportunities quantified by the volume of circumstances. 
 
Thus, a quantum system is formed. These five environmental properties takes physics to a new behaviour. The whole unity expressed by the symmetry of difference leads physics to evolve in time through set transformations. Introducing growth, evolution, emergence, complexity. Growth from potentiality and nonlinearity. Evolution means that nature allows on quantum individuation through growth with chance. Emergence is evolution under circumstance. Complexity is everything entrelaced.

\section{Variety physics}
${}$\indent

Historically, physics has being an atomist subject looking for the greek atm. Nevertheless conglomerates lead physics to the whole meaning. Nature develops systems and patterns resulting another type of individuation. The so-called third quantum. The quantum inserted in the whole.

Physics is not only the search for smaller and smaller things. Something together happens. Quantum systems states physics. Organizations follow under the antireductionist principle. Physical laws are more than ultimate constituents interactions. There is the diverse and how nature produces itself. Phase transitions show on laws depending on collective groupings. 

The constructivism introduces a set action physics to be investigated. The variety physics. Conglomerates form units generating own physics. New laws are constituted. A protagonism beyond reductionist subunits. Diversely from atomist physics based on interaction matter contains different levels of association to be considered. A systemic behaviour under interrelationship.

Nature performs varieties. A physics will environmental behaviour. Entities acting in grouping provide a second type of individuation. For instance, hydrogen and oxygen gases establishing water are showing that, isolated structures are diverse from grouping behaviour. Water is a variety derived from basic constituents with another properties. 

Physics is more than greek atms. Varieties introduces a way for understanding nature oppositedely than to break into little pieces. Differently from usual reductionist subunits, objects together are more. The novelty is that ultimate constituents does not work just as matryoshkas. Varieties in environment are constituing another type of elementarity. Nature also contains a determination derived from the fields set elementarities to be studied.

Varieties happens in all nature places. The elementarity related to a conglomerate. Proposing ultimate constituents not as lego but as parts in environment. Nature works out with a whole elementarity derived from associations. Physics and biology show conglomerates by composing matter and species. Generating differentiated structures with own character. Biology with genes individualizing species. The physics counter part should be preons and strings individualizing objects as varieties at eightfoldway, molecules and so forth. 

A qualitative change happens. The existence of antireductionist laws due to varieties. A collective physics where such elementarities and not ultimate constituents are responsible for another nature process. A systemic behaviour on set, diversity, interdependence, nonlinear, chance is expressed. A set dynamics is accomplished. The new physicality is that the whole unity is that evolves in time. Grouping physics under set transformations generates the variety dynamics. Determining on nature growth, evolution, emergence, complexity.

Evolution appears as a branch of the physics of varieties. Nature evolution is processed not through ultimate constituents but under the environment elementarities. The prescription of evolution was given by darwin. A determination under environment diversity, growth, chance, adaptaion. However, current physics is a research restricted to determine bodies under space-time dynamics. It does not face on mutation. Particles are not considered as corps in transformations. The quantum physics at $20^{th}$ century was limited to search for ultimate particles and their interactions. The relationship expressed between ultimate constituent, environment, variety and evolution still has to be constituted.

Complexity is another set consequence derived from the physics of varieties. A process where properties are depending on context. It is not described by fundamental subunits, but through cohesive sets. The reductionist approach does not work anymore. A multitude of facts integrated in a set action has to be considered. Quantum with complex synapses are expected.

Under this context, physics has to write down the mathematics of variety. New systemic properties about the whole quantum have to be expressed analytically. Gauge symmetry reinterpreted under antireducionist content propitiates a formalism. By studying the abelian constructivist lagrangian, one gets the existence of whole equations. A number of equations which have to be interpreted in a certain way. The grouping physics and corresponding set action express mathematically antireductionist properties as diversity, nonlinearity, interdependence, chance. They will be leading to set transformations, as evolution.

The variety mathematical expression is a study to follow. Eq. (\ref{equation 3.32}) contains a formulation to be expanded. Its proposal is to look analytically for the set action. Determine the relationship between the part and the whole. Describe a quantum system through constructivist equations. Express mathematically the presence of a third quantum.

\section{Set physics}
${}$\indent  

A set action is proposed by inserting a given field $G_{\mu I}$ in the fields set $\{G_{\mu I}\}$. Eqs. (2.6) and (3.29) are composing a constructivist physics based on an abelian antireductionist gauge symmetry. A third type of quantum is derived. An energy packet with antireductionist properties is composed.

A third quantum is being studied. It makes the atomist physics move for grouping physics. Five grouping properties will promote a whole quantum in network. They are set, diversity, interconnectivity, nonlinearity, chance. Introducing whole-parts relationships expecting to be expressed mathematized through the antireductionist gauge symmetry. A physics to be done.

The model contains the internal consistencies of locality, positive Hamiltonian, renormalizability and definite unitarity [20]. Given that, one can analyse the correspondent physical manifestations of such fields set action. The presence of an association with health quantum numbers.

\section{Diversity physics}
${}$\indent

The second grouping physics property is diversity. Analyse how far eq. (2.6) contains a set action under diversity. A symmetry establishing a pluriformity of diversities. Consider that, from a single gauge parameter, it is possible to gather particles with different quantum numbers.  

The thesis is that the set action provides a symmetry of diference. Consider that the quanta inserted in the abelian fields set are processing differences. The respective quantum numbers are multiple differentiated. They constitute a quantum system where spin, mass, charges, discrete symmetries are under a diversity. A quanta variety to be demonstrated at this section.

\subsection{Spin diversity}
\subsubsection{Sectors of spin-1 e spin-O}
${}$\indent

Given the field $A_{\mu I}$ belong to the representation of the Lorentz group $(j, k)=\left(\frac{1}{2},\frac{1}{2}\right)$, of the expression $s = [j - k]...(j+ k)$ we obtain the information that this representation contains two spin-1 and spin-0. Therefore, it is deduced that the set$\{A_{\mu I}\}$ contains  N-vector particles and $(N-1)$-scalars (one of which must be eliminated by gauge fixing) [21].

For understanding the variety of spins predicted from Lorentz group, one should analyze the Lagrangian. Rewriting L(G) in terms of its antisymmetric and symmetric sectors, we have:

\begin{equation}
    \lag (G) = \lag_A+\lag_S+\lag_{st}\label{antissimetric lagrangian}
\end{equation}
where
 
\begin{equation}
    \lag_A=b_Ib_JG_{\mu\nu}^IG^{\mu\nu J}+2b_I e_{[JK]} G_{\mu\nu}^I G^{\mu J} G^{\nu K}+e_{[IJ]}e_{[KL]} G_{\mu}^I G_{\nu}^J G_{\mu K} G_{\nu L},
\end{equation}

\begin{eqnarray}
    \no
    \lag_S&=&(\beta_I \beta_J) S_{\mu\nu}^I S^{\mu\nu J} + (2\beta_I\rho_J+4\rho_I\rho_J)S^{\mu I}_\mu S^{\nu J}_\nu + (2\beta_I e_{(JK)}) S_{\mu \nu}^I G^{\mu J} G^{\nu K}+\\
    \no
    &+&(2\beta_I e_{(JK)}) S_{\mu \nu}^I G^{\mu J} G^{\nu K}+ (2\beta_I \tau_{(JK)}+2\rho_I e_{(JK)}+8\rho_{I}\tau_{(JK)})S_\mu^{\mu I}S_\nu^{J}S^{\nu K}+\\\no
    &+& (e_{(IK)}e_{(JL)}+2e_{(IJ)}\tau_{(KL)}+4\tau_{(IJ)}\tau_{(KL)})G_\mu^IG^{\mu J}G_\nu^K G^{\nu L}
\end{eqnarray}
and

\begin{equation}
    \lag_{st}=\eta\epsilon_{\mu\nu\rho\sigma}(2b_I\mathbf{e}_{[JK]}G^{\mu \nu I} G^{\rho J} G^{\sigma k}+\mathbf{e}_{[IJ]}\mathbf{e}_{[KL]}G^{\mu I} G^{\nu J} G^{\rho K}G^{\sigma L})
\end{equation}

Nevertheless eq (\ref{antissimetric lagrangian}) show a model mixing spin-1 and spin-0. While $L_A$ carries spin-l, the symmetric and semi-topological sectors are mixing vectors and scalars. Therefore, in order to obtain a spectroscopy chemically pure, we must separate the kinetic term through transverse and longitudinal parts. The spins sectors involved in $\lag(G)$ as

\begin{equation}\label{L(G)}
    \lag(G)=\lag_T+ \lag_L +\lag_I
\end{equation}

The spin-1 sector given by

\begin{equation}\label{L_T}
    \lag_T=G_{\mu\nu I}G^{\mu\nu I}
\end{equation}

The spin-0 sector by
\begin{equation}\label{L_L}
    L_L=\beta_I\beta_J S_\mu^{\mu I} S_\nu^{\nu J} + \beta_I S_{\mu \nu}^{I}(\rho_J g^{\mu\nu}S_\rho^{\rho J})+\rho_Ig_{\mu\nu}S_\rho^{\rho I}(\beta_J S^{\mu \nu J}+\rho_J g^{\mu\nu}S_\sigma^{\sigma J})
\end{equation}
and the interaction term as
\begin{equation}\label{lag_I}
    \lag_I=\lag_I^A+\lag_I^S+\lag_I^{ST}
\end{equation}
where
\begin{eqnarray}\label{lag_I^A}
    &&\lag_I^A=(b_IG_{\mu\nu}^I+Z_{[\mu\nu]})\mathbf{e}_{[\mu\nu]}\\
    &&\lag_I^S=(\beta_IS_{\mu\nu}^I+\rho_Ig_{\mu\nu}S_\rho^{\rho I} + Z_{(\mu\nu)})(\mathbf{e}^{(\mu\nu)}+g^{\mu\nu}\omega_\rho^\rho)\\
    \no&&\lag_I^{ST}=\eta\epsilon_{\mu\nu\rho\sigma}(2b_I\mathbf{e}_{[JK]}G^{\mu \nu I} G^{\rho J} G^{\sigma k}+\mathbf{e}_{[IJ]}\mathbf{e}_{[KL]}G^{\mu I} G^{\nu J} G^{\rho K}G^{\sigma L})\\
\end{eqnarray}

Thus, eq. (\ref{L(G)}) decants the first quantum number associated at the set $\{G_{\mu I}\}$. Since spin is the most direct physical entity provided by field theory, eqs.(\ref{L_T} e \ref{L_L}) respectively separate spin-1 and spin-0 sectors. Consequently, with defined spins, we will be able to follow about other quantum numbers being proposed by this systemic model [22].

A further discussion is how far one can separate the spin sectors without violating localizability. Decomposing $A_{\mu I} \equiv A_{\mu I}^{T} + A_{\mu I}^{L}$ where $A_{\mu}^{T} = \theta_{\mu \nu}A^{\nu}$, $A_{\mu}^{L} = \omega_{\mu \nu}A^{\nu}$ and considering the Lorentz transformation $x'_{\mu} = \Lambda_{\mu}^{\nu}x_{\nu}$, one gets the relationships

\begin{eqnarray}
    (A'_{T})_\mu = (\Lambda^{-1})^{\rho}_\mu (A_{T})_{\rho} \text{;  }(A'_{L})_\mu = (\Lambda^{-1})^{\rho}_\mu (A_{L})_{\rho}\label{Transformation for sectors}
\end{eqnarray}

Eq.(\ref{Transformation for sectors}) is showing that the fields transverse and longitudinal decompositions are consistent with Lorentz covariance. This guarantee these two sectors under different Klein Gordon equations without violating Poincarè. They will carry different dynamics preserving the original symmetries. 

\subsection{Mass diversity}
\indent${}$

Mass is the next quantum number to be understood. Considering the spin sectors, and given the possibility two separate them consistently, next will be to the understand their masses constitutions. Split in transverse and longitudinal masses.

Physical masses are obtained by decomposing the kinetic part in terms of transverse and longitudinal projectors. According to Appendix A, it yields
\begin{equation}
    L_K=L_K^T+L_K^L=\frac{1}{2}G^t_\mu (\Box+m^2_T)\theta^{\mu\nu}G_\nu +\frac{1}{2} G^t_\mu(\tilde{B}\Box+m^2_T)\omega^{\mu\nu}G_\nu
\end{equation}
where $G^t_\mu=(G_{\mu 1},\dots,G_{\mu n})$, $\theta_{\mu\nu}-\frac{\partial_\mu\partial_\nu}{\Box}$ e $ \omega_{\mu\nu}=\eta_{\mu \nu}\frac{\partial_\mu\partial_\nu}{\Box}$.

The abelian symmetry enlargement is introduced. While the standard atomist gauge theory states that the number of gauge bosons is determined by the number of group theory generators, a variety of quanta is expressed at eq. (8.12).  Analizying the correspondent spin-1 and spin-0 sectors, one gets masses $m^{2}_{T}$ and $m^{2}_{L}$. Apendix A is showing transverse $m^{2}_{T}$ and longitudinal masses $m^{2}_{L}$ as eigenvalues of matrizes $K^{-1}_{T}M^2$ and $(K_{L}+G_{F})^{-1}M^{2}$ respectively.

\subsubsection{Transverse masses}
${}$\indent

As stipulated by the $\Omega$ matriz transverse diagonalization, it fixes the physical fields $\{G_{\mu I}\}$ associated with spin-1 and corresponding physical mass, $m^2_T$. The transverse propagates are expressed as $<G_{\mu I} G_{\nu J}> = \frac{i\delta_{IJ}}{\kappa^{2}-m^{2}_{T, IJ}}$. It yields that $\lag_T(G)$ contains a diagonalized mass matrix $m^2_T$ given by

\begin{equation}
    m^2_{T}=\begin{pmatrix}
        0 & 0 & ... & 0\\
        0 & m^2_{T, 22} & ... &0\\
        \vdots&\vdots& &\vdots\\
        0&0&\dots&m^2_{T, NN}
    \end{pmatrix}
\end{equation}
Eq. (8.13) shows the mass spectroscopy corresponding to  spin-1 sector. Notice that the system necessarily contains a mandatory zero mass and others whose value depends on free coefficients analyzed at section 4.

The physical choice was to make the transverse sector completely diagonalized. The relationships $M^{2}=(\Omega^{2})^{-1}m^{2}\Omega^{-1}$ and  $\Omega^{t}K_{T}\ =\mathbb{I}$ yields that $\Omega \Omega^{t} = K^{-1}_{T}$, which gives $K_{T}^{-1}m^{2}=\Omega m \Omega^{-1}$. This expression says that the matrices that contain the poles in both parametrization systems, $\{D_{\mu},X_{\mu}^{i}\}$ and $\{G_{\mu \pm}\}$ are related by a similarity transformation. A result showing the poles invariance under fields transformation. Another possibility would be diagonalize the longitudinal sector. Apendice -- works on that.

\subsection{Longitudinal masses}
${}$\indent
We must now discuss the question of the poles in the L-sector of the field propagators $G_{\mu I}$. These are given at expression $(\tilde{B} \Box+ m^2_{T})^{-1}\omega_{\mu\nu}$ where $\Tilde{B}=\Omega^{t}(K_{L}+G_{F})\Omega$. Therefore, we must ensure that the matrix $(\tilde{B}\Box+ m^2_{T})$ is invertible. However, $m^2_{T}$ can have one or more zero eigenvalues, so it is not, in general, invertible. So, knowing that 'the sum of two matrices is invertible, if and only if one of the matrices is invertible,' we have that, the invertibility of $(\tilde{B} \Box+ m^2_{T})$ is due, in general, to the invertibility of $\tilde{B}$. Thus, returning to the propagators at L-sector, we obtain

 \begin{equation}
     <G_{\mu I}G_{\nu J}>_{L}=\left[ \frac{i}{\Box+\tilde{B}^{-1}m^2_{T}}\tilde{B}^{-1}\right]_{IJ}\omega_{\mu\nu}
 \end{equation}
resulting that the poles of the longitudinal sector are obtained as the eigenvalues of $\tilde{B}^{-1}m^2_T$. It should be noted that the $\tilde{B}$ matrix also contains the gauge-fixing parameters, so in principle its poles should be bringing a dependency on the gauge parameter. It yields (N-1) quanta where one is suppressed by the gauge fixing term.

The following step to analyze is that the T-sectors and L-sectors are not completely independent. There is a relationship between the masses of both sectors given by the expression $det(\tilde{B}^{-1} m^2_T) = \frac{det\text{ }m^2_T}{det\text{ }\tilde {B}}$, which results, 

\begin{equation}
    (m^2_1\times m^2_2\times\dots m^2_N)=det\tilde{B}\cdot det(\tilde{B}^{-1}m^2_T)
\end{equation}

Therefore, given $det \tilde{B} \neq 0$ due to $\tilde{B}$ being invertible, eq. (7.15) is that given that necessarily $m^2_{T,11} = 0$, we have

\begin{equation}
    det(\tilde{B}^{-1}m^2_T)=0=det(m^2_L)
\end{equation}

Thus, a question is the degenerescence degree correspondence between a null $m_{T}^{2}$ and $m_{L}^{2}$ eigenvalues. For understanding, consider at sector T, that between N-fields $G_{\mu I}$ with M independent fields $(M\leq N)$ containing zero mass. Considering that vectors $G_{\mu 1},...,G_{\mu N}$ are also independent eigenvectors of $(\Tilde{B}^{-1})^{t}m^{2}$, one gets

\begin{eqnarray}\label{massa longitudinal}
    m_{L}^{2}G_{\mu 1}=0, ..., m_{L}^{2}G_{\mu M}=0
\end{eqnarray}
where eq. (\ref{massa longitudinal}) shows that the degenerescence degree for null $m_{T}^{2}$ eigenvalues.

Thus eq (7.15-7.17) are expressions to characterize mass spectroscopy. Saying that, if there is a zero mass in the T-sector, it will correspond to another zero in the L-sector. Physically, a zero-mass spin-1 photon will be associated with another zero-mass scalar photon.

The next step to ensure the physicality of the model is that the physical masses $m^2_T$ and $m^2_L$ of the transversal and longitudinal sectors do not contain tachyons. Control the relations $m^2_T. \ge 0, m^2_L \ge 0$ through the free coefficients of the theory. These relationships will be controlled by the free coefficients of theory.

\subsection{Charges diversity.}
${}$\indent

The set action produces three types of charges. First, corresponding to the 2N-equations of motion derived from eqs. (8.5-8.6). Second, the Noether charge associated to gauge symmetry. Third, that one related to the symmetric Bianchi identity given by eq. (10.3).

A physics beyond electric charge is developed. It shows, that from a common abelian symmetry it is possible to generate $(2N+2)$ conserved charges. They are differentiated but with interdependent nature given by the fields set $\{G_{\mu I}\}$. An elecromagnetism with conservation laws beyond electric charge is derived to be analysed.

Thus, the N-vector field system $\{G_{\mu I}\}$ constitutes $(2N +2)$ conserved charges. The basic charge term is potential fields interacting with fields strengths. They are observed at equations of motion, Noether, Bianchi identities. Their expressions make diverse compositions of the $G_{\mu I}$ fields and fields strengths with granular, collective antisymmetric, symmetric nature. It yields abelian fields selfinteractions working as sources for fields charges.

\subsection{Discrete symmetries}
${}$\indent
The next study on quantum diversity through the set notion is to consider the discrete symmetries [23].

\subsubsection{P, C, T and PCT symmetries}
${}$\indent

Given the fields set $\{G_{\mu I}\}$, one has to understand on the possibilities of allocating a diversity of discrete symmetries. Starting from $L(G)$ seen by eq. (3.7),

\begin{eqnarray}
    L=&\frac{1}{2}&G_\mu^t(\Box+m^2_{T})\theta^{\mu\nu}G_\nu+\frac{1}{2}G_\mu^t(\tilde{B}\Box+m^2_{T})\omega^{\mu\nu}G_\nu\nonumber
    \\    &+&\partial_{\mu}G_\nu^t\tilde{\lambda}G^{\mu}G^{\nu}+G^{t}_{\mu}G^{t}_{\nu}\tilde{\Lambda}G^{\mu}G^{\nu}
\end{eqnarray}

\noindent where $G_{\mu}^{t}=(G_{\mu 1},\dots,G_{\mu N})$ and $m^2_{T}=diag(m_{11}^2=0,\dots,m_{NN}^2)$.

The purpose is to observe the discreted symmetries P, C, T for diverse whole quanta [23], without breaking the PCT structure characterized by the theorem due to Luders, Pauli and Schwinger [24].

Starting from parity [25],

\begin{equation}
    G_{\mu I}(x) \stackrel{P}{\to} UG_{\mu I}(x)U^{-1}=G'_{\mu I}(x')=\eta_P G_{I}^{\mu}(x)
\end{equation}

\noindent where $t'=t,\vec{x'} = -\Vec{x}$ end $(\eta_P)_{IJ}=\eta_P\delta_{IJ}$ is the intrinsic parity. For, $\eta_P = 1$, implies vector, $\eta_P= - 1$, pseudo-vector. Replacing (7.19) in eq. (7.18), we obtain 

\begin{eqnarray}
    &&L^{'}=\frac{1}{2}G_{\mu}^{t}(\Box+m^{2}_{T})\theta^{\mu \nu}G_{\nu}+\frac{1}{2}G_{\mu}^{t}(\eta_{P}\Tilde{B}\eta_{P}+m^{2}_{T})\omega^{\mu \nu}G_{\nu}\nonumber
    \\
    &&\partial_{\mu}G_{\nu}^{t}\eta_{P}\Tilde{\lambda}\eta_{P}\eta_{P}G_{\mu} G_{\nu}+G_{\mu t}G_{\nu t}\eta_{P}\eta_{P}\Tilde{\lambda}\eta_{P}\eta_{P}G_{\mu}G_{\nu}
\end{eqnarray}
which yields the following conditions for P-invariance:

 \begin{align}
\no&\eta_P\tilde{B}\eta_P=\tilde{B} &  &\text{ or } &  &\tilde{B}_{IJ} = \eta_{IM}\tilde{B}_{MN}\eta_{NJ}\\
\no&\eta_P\tilde{\lambda}\eta_P\eta_P=\tilde{\lambda} &  &\text{ or } &  &\tilde{\lambda}_{IJK} = \eta_{IM}\tilde{\lambda}_{MNP}\eta_{MJ}\eta_{PQ}\\ 
&\eta_P\eta_P\tilde{\Lambda}\eta_P\eta_P=\tilde{\Lambda} &  &\text{ or } &  &\tilde{\lambda}_{IJKL} = \eta_{IM}\eta_{JN}\tilde{\Lambda}_{MNPQ}\eta_{PQ}\eta_{LQ}
\end{align}

Equity (8.21) shows that parity is naturally conserved when all $G_{\mu I}'$ are vectors $(\eta_{IM}=-\delta_{IM})$. However, for pseudo-vectors $(\eta_{IM}=\delta_{IM})$, the trilinear term is not satisfied. For mixed case, one should study the restrictions of eq. (7.21). For example, for the case of three fields with two vectors $(A'_{\mu}(x')=A_{\mu}(x))$ and a pseudo-vector $(A'_{\mu}(x')=-A_{\mu}(x))$, $\eta$ $\eta_p= diag(1, 1, -1)$. For three linear term one must explicitly derive the relationships between the coefficients of the theory. For four linear terms both cases are directive preserved.

The second discrete symmetry is charge conjugation [26]. It reverses the signs of all non-zero generalized charges of a given particle, transforming it into the corresponding antiparticle. Considering complex fields $G_{\mu I}$,

\begin{equation}
    G_{\mu I}(x) \stackrel{C}{\to} U_C G_{\mu I}(x)U_C^{-1}=G'_{\mu I}(x')=\eta_CG_{\mu I}^{*}(x)
\end{equation}

The transverse kinetic term is automatically invariant, $\eta_C^2 = 1$ and $\eta_c m^2_{T} \eta_c = m^2_{T}$. But, for other terms in $L$ one must establish the corresponding conditions. Substituting (8.22) into (8.13), we get 

\begin{eqnarray}
    L\stackrel{C}{\to}L'=L 
\end{eqnarray}
for
 \begin{align}
&\tilde{B} = \eta_C\tilde{B}\eta_C, &  &\tilde{\lambda} = \eta_C\tilde{\lambda}\eta_C\eta_C&  &\tilde{\Lambda} = \eta_C\eta_C\tilde{\Lambda}\eta_C\eta_C
\end{align}

Equity (7.25) shows that a C-invariance is obtained when all $G_{\mu I}'s$ have $\eta_C=1$. In the case when the $G_{\mu I}'s$ have $\eta_C=-1$ (photonic case), we have $\lambda = 0$, which means that the Lagrangian does not consider trilinear interactions for the same field, as prescribed by Furry's theorem [27].

In this context, it is observed that L and separately the fields can break the parity and charge conjugation symmetries. However, it is also possible that the combined operation PC is invariant. 

The third case is the anti-unit operation of time inversion [28],

\begin{equation}
    G_{\mu I}(x) \stackrel{T}{\to} U_T G_{\mu I}(x)U_T^{-1}=G'_{\mu I}(x')=\eta_TG_{\mu I}(x)
\end{equation}

\noindent where $x'^\mu=x^\mu,\partial'_\mu=\partial_\mu,\Box'=-\Box, \theta'_{\mu\nu}=\theta_{\mu\nu},\omega'_{\mu\nu}=\omega_{\mu\nu}$ 

Considering that $m^2_{T},\tilde{B}, \tilde{\lambda}$ and $\tilde{\Lambda}$ are real matrices ($\Omega$ is real), the transverse term is naturally invariant while the longitudinal and interaction sectors depend on the relations

\begin{align}
    &\tilde{B} = \eta_T\tilde{B}\eta_T, &  &\tilde{\lambda} = -\eta_T\tilde{\lambda}\eta_T\eta_T&  &\tilde{\Lambda} = \eta_T\eta_T\tilde{\Lambda}\eta_T\eta_T
\end{align}

\noindent where eq. (7.25) means that trilinear terms are prohibited for all $G_{\mu I}'s$ with $\eta_T=1$. In the case where all of them have $\eta_T=-1$, the T-invariance is automatic. 

The operation PCT should be performed for proofing the model consistency. Based on the so-called PCT theorem, due to Lürders, Pauli and Schwinger,which states that the product of the transformations P,C,T considered in any order, is always a symmetry of a quantum theory, when the Lagrangian that defines it is: local, hermitian, invariant under proper Lorentz transformation and if the fields are quantized in accord with the usual spin-statistics correction. Now, one has to verify how this PCT operation acts for a whole system.

The PCT symmetry is given by

\begin{equation}
    G_{\mu I}(x) \stackrel{PCT}{\to} U_TU_CU_P G_{\mu I}(x)U_T^{-1}U_C^{-1}U_P^{-1}=G'_{\mu I}(x')=\eta_{PCT}G^{*}_{\mu I}(x)
\end{equation}
where $\eta_{PCT}=\eta_P\eta_C\eta_T$ with $x'^\mu=x^\mu, \partial'_\mu=\partial_\mu$

\begin{eqnarray}
    L\stackrel{PCT}{\to}L'&&=\frac{1}{2}G_{\mu}^{t}\eta_{PCT}(\Box+m_{T}^{2})\theta^{\mu \nu}\eta_{PCT}G_{\nu}+\nonumber
    \\
    &&+\frac{1}{2}G_{\mu}^{t}\eta_{PCT}(\Tilde{B}\Box+m_{T}^{2})\omega^{\mu \nu}\eta_{PCT}G_{\nu}+\nonumber
    \\
&&+\partial_{\mu}G_{\nu}^{t}\eta_{PCT}\Tilde{\xi}\eta_{PCT}G^{\mu}\eta_{PCT}G^{\nu}+\nonumber
    \\
&&+G_{\mu}^{t}\eta_{PCT}G_{\nu}^{t}\eta_{PCT}\Tilde{\Lambda}\eta_{PCT}G^{\mu}\eta_{PCT}G^{\nu}
\end{eqnarray}  

Thus, the transverse term is automatically invariant ($\eta_{PCT}^2=1,\eta_{PCT}m^2_{T}\eta_{PCT}=m^2_{T}$) Meanwhile, the other terms require the following conditions:

\begin{eqnarray}
    \no&&\eta_{PCT}\tilde{B}\eta_{PCT}=\tilde{B}\\
    \no&&\eta_{PCT}\tilde{\lambda} \eta_{PCT}\eta_{PCT}=-\tilde{\lambda}\\
    &&\eta_{PCT}\eta_{PCT} \tilde{\Lambda}\eta_{PCT}\eta_{PCT}=\tilde{\Lambda}
\end{eqnarray}

Consequently the transverse term is invariant, but the longitudinal depends on eqs. (8.29). However, as we know, PCT theorem requires a closure not related to theory free coefficients. Then, in order to express it in a more generic way, not depending on these coefficients, we are going to take $V_{\mu}$-referential system. From Eq. (10) and $G_{\mu}=\Omega^{-1}V_{\mu}$,

\begin{eqnarray}
    V_{\mu}(x) \stackrel{PCT}{\to} U_{PCT}V_{\mu}(x)U_{PCT}^{-1} = V'_{\mu} = \mathcal{N}V_{\mu}(x),
\end{eqnarray}
where $\mathcal{N} \equiv \Omega\eta_{PCT}\Omega^{-1}$. Substituing Eq. (12) in Eq. (2), one derives the following relationships: $\mathcal{N}^{t}K\mathcal{N}=K$ and $\mathcal{N}^{t}M^{2}\mathcal{N}=m^{2}$ showing that transverse part is PCT invariant. Similarly the longitudinal part is invariant through $\mathcal{N}^{t}B\mathcal{N}=B$, trilinear term for $\mathcal{N}^{t}\lambda \mathcal{N}\mathcal{N}=- \lambda$ and quadrilinear for $\mathcal{N}^{t}\mathcal{N}^{t}\Lambda \mathcal{N}\mathcal{N}=- \Lambda$. These relationships show that PCT teorem is explicity verified reference $V_{\mu}$. Consequently, transporting the PCT-invariance at reference $V_{\mu}$ at reference $G_{\mu I}$, it provides PCT-invariance not depending on any freecoefficient. 

Thus, PCT theorem is verified. As expected it depends only on Lagrangian nature not on its expression. In summary, $\lag \stackrel{PCT}{\to} \lag' = \lag$, and a consistency for this whole abelian model is obtained.

Considering that$ L(G)$ is local and hermitian the CPT invariance should obtained as a whole [24]. Nevertheless, the appearance of these invariance conditions is yet to be understood. It was expected  that given the invariance conditions under Poincaré, locality and hermitity for an L means that it should automatically respect CPT.

The above equations are showing that the discrete symmetries of the transverse sector are regulated by the coefficients $\eta_p,\eta_c,\eta_T$ while those of the longitudinal sector are regulated by the coefficients of the matrix $\tilde{B}$. Allowing a given $G_{\mu I}$ field to accommodate different discrete symmetries for its vector and scalar quanta. In this scenario, the $\{G_{\mu I}\}$ set contains groups of spin-1 and spin-0 quanta featuring a variety of discrete symmetries.

\subsection{Symmetry of difference}
${}$\indent

At previous sections, a constructivist physics was proposed. We are assuming the nature set action. The relationship between the part and the whole introduced a third quantum. It includes another a physics perspective. Quanta diversity and antireductionist properties are new aspects to include at quantum physics. Two features requiring to be covered by a symmetry. It will be called as the symmetry of difference.

Eq. (2.6) supports a new symmetry. The atomism is replaced by constructivism. A physics between set and diversity is constituted. The antireductionist gauge symmetry proposes differences acting together. A new quantum creation is introduced. Gauge symmetry with the sense of unity at difference. A unity producing a conglomerate physics with own properties.

The symmetry of difference is introduced by constructivist Lagrangian. By analyzing the quanta spectroscopy at set $\{G_{\mu I}\}$, a diversity is observed. As studied, the quantum numbers corresponding to each quanta embedded in the fields set $\{G_{\mu I}\}$ are differentiated. The correspondent spin, mass, charges; P, C, T are under a diversity without violating the gauge symmetry and PCT theorem. Consequently, the third quantum existence is not only based on the part inserted in the whole, but under quantum diversity. 

A quantum system is established under the unity of difference. The set action $\{G_{\mu I}\}$ contains multiple quanta. The quanta pluriformity. A whole with differentiated parts. A new elementarity is proposed. Quantum systems with varieties. Differentiated quanta and whole physical properties are established by the antireductionist gauge symmetry. The meaning of variety is established. There is a new way to do physics. Instead of wave function consider variety. A construction where the fields set action differentiates parts under gauge symmetry. A  quantum system is constituted by interconnected fields.
 
Three achievements are obtained. First, the symmetry of difference replace the spontaneous symmetry breaking and grand unification. Second, it yields a third quantum physics. The fields set action interrelates equations of motion, Bianchi identities, conservation laws, Lorentz forces. Third, quantum systems are redefined by substituting quantum mechanics through gauge symmetry.

The set action turns out a new physical performance to be studied. The symmetry of difference constitutes an alternative to grand unification and spontaneous symmetry breaking. It minimizes the physical conditions to carry physical entities. It does not follow the group theory rule to gauge theory established by the Göttingen group since 1920s.

The symmetry of difference breaks the symmetry relationship established by group theory. Two statements were constituted. On the degeneracy of physical states and the number of intermediate bosons. The quantum mechanics development in the 1920s created a standard for the use of symmetry in physics. To assert whether G is the symmetry group of a quantum system, the states of this system are grouped into representations of G called degeneracies. For example, according to the SO(4) group, the hydrogen atom (quantum system) has $2N^{2}$ degenerate states where N is the energy level [29]. 

In particle physics, instead of degenerate states, the presence of particles is introduced. And so, from the point of view of relativity, $E=mc^{2}$, degeneracy is characterized by the existence of particles with the same mass. This proposal was considered in 1932 by Heisenberg when he included the proton and the neutron (mass difference $0.1\%$) in a some multiplet and established the isospin symmetry. This idea was followed by Yang-Mills in 1954 where the number of associated gauge bosons must be equal to the number of group generators [12-13]. For instance, SU(5) requires 24 intermediated particles [30]. It also simplifies on mass introduction. A second aspect of the symmetry of difference is that it may be manifested without implemeting an ad hoc a scalar potential. Higgs fields are not necessary [30].

\subsubsection{Example: symmetry of difference for a 
    quantum system with five bosons}
${}$\indent

Quantum system featuring a quantum diversity must be exemplified. Consider five $G_{\mu I}$ fields constituting the set $\{G_{\mu 1}, G_{\mu 2}, G_{\mu 3}, G_{\mu 4}, G_{\mu 5}\}$. To examine the inclusion of quantum numbers differentiations with respect to different masses, coupling constants and invariances C, P, T, and CPT, let us consider the following possibilities:

\begin{enumerate}[(i)]
    \item $J^{PC}=1^{--}$, like the photon, $J/\psi$, $\rho(770)$. Identify such particles as $G_{\mu 1},G_{\mu 2}, G_{\mu 3}$.    
    \item $J^{PC}=1^{+-}$, such as $b_1(1235)$. associate particle as $G_{\mu 4}$.
    \item $J^{PC}=1^{++}$, like the $X_1(3510)$. Identify the particle as $G_{\mu 5}$.
\end{enumerate}

Thus, $L(G)$ at eq. (8.2) accomodates the spin-1 and spin-0 sector with different masses and coupling constants. Then, we are going study on discrete symmetry. In addition, to the usual discrete individual symmetries of each $G_{\mu I}$ field, there are discrete systemic symmetries for grouping $\{G_{\mu I}\}$. The perspective what counts is the symmetry of $L(G)$. Physically, discrete symmetry can act differently for each $G_{\mu I}$ field interaction. Implying that, there is the set invariance and a diversity of individual discrete symmetries.

\indent Consequently, the set $\{G_{\mu },G_{\mu 2},G_{\mu 3},G_{\mu 4},G_{\mu 5}\}$ formed by the composition of these particles will preserve CPT, but promoting interactions which preserve or not the individual discrete symmetries. Discrete symmeties may considered individualy or as a whole $\{G_{\mu },G_{\mu 2},G_{\mu 3},G_{\mu 4},G_{\mu 5}\}$. For example, as a whole it is invariant under parity, although individually several of its fields are not.

\section{Nonlinear physics}
${}$\indent
A whole quantum physics is being developed. The usual relativistic fields equations are developed in a fields set minimal action, $\delta S = 0$. It gives nonlinear a dynamics constituted by coupled equations of motion [31].

\begin{eqnarray}
    &&\partial_{\mu}\{(G^{\mu \nu}_{I} + a_{I}\mathbf{e}^{[\mu \nu]}) + \eta \epsilon^{\mu \nu \rho \sigma}(G_{\rho \sigma I} + a_{I}\mathbf{e}_{[\rho \sigma]})+\nonumber
    \\
    &&+\beta_{I}(\beta_{J}S^{\mu \nu J} +\rho_{J}g^{\mu \nu}S^{\alpha J}_{\alpha} + \mathbf{e}^{(\mu \nu)} + g^{\mu \nu}\mathbf{\omega}^{\alpha}_{\alpha})\}+\nonumber
    \\
    &&+\partial^{\nu}\{(\beta_{I}\beta_{J}+\beta_{I}\rho_{J}+\beta_{J}\rho_{J}+\frac{1}{4}\xi_{IJ})S^{\alpha J}_{\alpha}+\rho_{I}(\mathbf{e}^{\alpha}_{\alpha}+4\omega^{\alpha}_{\alpha})\}\nonumber
    \\
    &&+ \frac{1}{2}\mathbf{m}_{II}^{2}G^{\nu I} = J^{\nu}_{I}(G)
\end{eqnarray}
where

\begin{eqnarray}
    J^{\nu}_{I}(G) = J^{A \nu}_{I}(G) + J^{S\nu}_{I}(G) + J^{st \nu}(G)\nonumber
\end{eqnarray}
are expressing a nonlinearity
\begin{eqnarray}
    &&J^{A\nu}_{I}(G) = \mathbf{e}_{[IJ]}G^{J}_{\mu}Z^{[\mu \nu]},\nonumber
    \\
    &&J^{S\nu}_{I}(G) = \mathbf{e}_{(IJ)}G_{\mu}^{J}Z^{(\mu \nu)} + \tau_{(IJ)}G^{\nu J}Z_{\alpha}^{\alpha}\nonumber
    \\
    &&\nonumber
    \\
    &&J^{st \nu}_{I}(G) = \eta\mathbf{e}_{[IJ]}\epsilon^{\mu \nu \rho \sigma}G_{\mu}^{J}Z_{[\rho \sigma]} + \eta b_{I}\epsilon^{\mu \nu \rho \sigma}\partial_{\nu}\mathbf{e}_{[\rho \sigma]}\label{equation 8.3}
\end{eqnarray}

Eq. (9.1) can be reduced by using the kinetic identity $\partial_{\mu}S^{\mu \nu}_{I} = \partial_{\mu}G^{\mu \nu}_{I}+\partial^{\nu}S^{\mu}_{\mu I}$. It will suppress the fields strenghts $S^{\mu \nu}_{I}$ and $e^{\mu \nu}$ from the dynamics. The spin-1 and spin-0 sectors are separated.

Eq. (\ref{equation 8.3}) are sources carrying the fields set $\{G_{\mu I}\}$. A selfprocess where fields strengths $Z_{\mu \nu}$ work as own sources. Through its own dynamics. $Z_{\mu \nu}$ is moving at space-time and growing. Developing a quantum  self creation own internal sources.

\section{Systemic physics}
${}$\indent

Considering eq. (2.6) where fields set $\{G_{\mu I}\}$ is sharing a common gauge symmetry, an interrelatioship is expected. A system working under a fields set unity. Fields are interwined by collective fields strengths and by equations of motion, Faraday laws, conservation laws, constructive Lorentz forces. It yields a systemic physics producing antireductionist interaction, induction, connection [32].

Separating eq. (9.1) in transverse and longitudinal sectors, it gives

For spin 1:
\begin{eqnarray}
    \partial_{\mu}\{G^{\mu \nu}_{I} + a_{I}\mathbf{e}^{[\mu \nu]}\}+\frac{1}{2}\mathbf{m}^{2}_{II}G^{\nu}_{I} = J^{\nu}_{I}(G)
\end{eqnarray}

For spin-0:

\begin{eqnarray}
    &&\partial^{\nu}\{(\beta_{I}\beta_{J} + \beta_{I}\rho_{J} + \rho_{I}\beta_{J}+\rho_{I}\rho_{J}+\xi_{IJ})S^{\alpha J}_{\alpha}+ (-\frac{1}{2}\beta_{I}+ \rho_{I})\mathbf{e}^{\alpha}_{\alpha} + \nonumber
    \\
    &&+(\beta_{I} + 4\rho_{I})\omega^{\alpha}_{\alpha}\}+\frac{1}{2}\mathbf{m}^{2}_{II}G^{\nu}_{I} = J^{\nu}(G)_{I}-\frac{1}{2}\mathbf{e}_{IJ}S^{\alpha}_{\alpha J}G^{\mu}_{I}\nonumber
    \\
    &&- \mathbf{e}_{(IJ)}S^{\nu \mu}_{J}G_{\nu I}
\end{eqnarray}
The above differential equations system explicit the $\{G_{\mu I}\}$ interrelationships.

The Bianchi identities improve the above systemic interrelationships. They provide the inductions below

\begin{eqnarray}
    &&\partial_{\mu}G^{I}_{\nu \rho} + \partial_{\nu}G^{I}_{\rho \mu} + \partial_{\rho}G_{\mu \nu}^{I} = 0\nonumber
    \\
    &&\partial_{\mu}\mathbf{e}_{[\nu \rho]} + \partial_{\nu}\mathbf{e}_{[\rho \mu]} + \partial_{\rho}\mathbf{e}_{[\mu \nu]} = \mathbf{e}_{[IJ]}G_{\nu}^{I}G^{J}_{\mu \rho} + \mathbf{e}_{[IJ]}G_{\mu}^{I}G^{J}_{\rho \nu} + \mathbf{e}_{[IJ]}G_{\rho}^{I}G^{J}_{\mu \nu} \nonumber
    \\
    &&\partial_{\mu}\mathbf{e}_{(\nu \rho)} + \partial_{\nu}\mathbf{e}_{(\rho \mu)} + \partial_{\rho}\mathbf{e}_{(\mu \nu)} = \mathbf{e}_{(IJ)}G_{\nu}^{I}S^{J}_{\mu \rho} + \mathbf{e}_{(IJ)}G_{\mu}^{I}S^{J}_{\rho \nu} + \mathbf{e}_{(IJ)}G_{\rho}^{I}S^{J}_{\mu \nu}\nonumber
    \\
    &&\partial_{\mu}\omega^{\nu}_{\nu} + 2\partial_{\nu}\omega_{\nu}^{\nu}=\tau_{(IJ)}G^{I}_{\mu}S^{\nu J}_{\nu} + 2\tau_{(IJ)}G^{I}_{\nu}S^{\mu J}_{\mu}\label{equation 9.3}
\end{eqnarray}

Eqs. (\ref{equation 9.3}) retreat the granular and collective, antisymmetric and collective symmetric Bianchi identities. It yields the following conservation law $\partial_{\mu}J^{\mu}_{B}=0$ where $J^{\mu}_{B}=\varepsilon^{\mu \nu \rho \sigma}e_{[IJ]}G^{I}_{\nu}G^{J}_{\rho \sigma}$.

The three Noether identities are

\begin{eqnarray}
    &&\partial_{\mu}J^{\mu}_{N} = 0 \nonumber
    \\
    &&\Omega_{I1}\partial_{\nu}\frac{\partial L}{\partial(\partial_{\nu}G_{\mu I})} + J^{\mu}_{N} =0 \nonumber
    \\
    &&\Omega_{I1}\frac{\partial L}{\partial(\partial_{\mu}G_{\nu I})}\partial_{\mu}\partial_{\nu}\alpha = 0
\end{eqnarray}

Considering the energy-momentum tensor, one gets another relationships through the conservation law [33]. 

\begin{eqnarray}
    &&\frac{\partial U}{\partial t} + \vec{\nabla}\cdot\vec{S} = 0,
    \\
    &&\frac{\partial S_{i}}{\partial t} - \partial_{j}T_{ij} = 0
\end{eqnarray}

The Lorentz constructivist force between the fields inserted at $\{G_{\mu I}\}$ is extended for the following grouping interrelationships [22]

\begin{eqnarray}
    f^{\nu} = f^{\nu}_{L} + f^{\nu}_{M} + f^{\nu}_{E}
\end{eqnarray}
with
\begin{eqnarray}
    f^{\nu}_{L} = 4G^{\nu \mu I}j_{\mu I}, \text{ } f^{\nu}_{M} = -2m_{I}^{2}G^{\nu}_{I}(\partial^{\mu}G_{\mu}^{I}), \text{ }f^{\nu}_{E} = 4 G^{\nu I}(\partial^{\mu}J_{\mu I}(G))
\end{eqnarray}
where $j_{\mu I}$ is an external source and $J_{\mu I}(G)$ defined at eqs (9.2). At vectorial shape,
\begin{eqnarray}
    &&\vec{f}_{L} = 4\rho_{I}\vec{E}^{I} + 4\vec{j}_{I}\times\vec{B}^{I},
    \\
    &&\vec{f}_{M} = -2m^{2}_{I}\vec{G}_{I}(\frac{\partial}{\partial t}\phi^{I} +\vec{\nabla}\cdot\vec{G}^{I}),
    \\
    &&\vec{f}_{E} =4\vec{G}_{I}(\frac{\partial}{\partial t}\rho^{I}(G) + \vec{\nabla}\cdot\vec{J}^{I}(G))
\end{eqnarray}

The above equations are showing how fields inserted in a fields set $\{G_{\mu I}\}$ are producing constructivist fields forces. Lorentz force is reproduced at eq. (10.9) and extended for a fields set. Eqs. (10.9-10.11) are introducing a force dependence on masses and fields. It yields a perspective different from usual atomist gauge theory which is just providing their interactions with external sources.

Summarizing, a quantum system is formulated mathematically. The above relationships between the whole and the part (whole quantum or variety) originated from eq. (2.6) produce a classical structure with $3N+11$ equations. These equations are showing a set physics. A systemic interdepent determinism where the individual, collective and whole are related together.

\section{Chance physics}
${}$\indent

A grouping physics is being derived. A new whole quantum significance proposed. Instead of ultimate constituents its definition comes out from the set. The quanta expressed from the fields set $\{G_{\mu I}\}$ introduce new perspectives. Next consider on the quantum chance. 

Let us start with the spin-1 sector. The matrix $K_{T}$ expressed at (C.8) can be separated into four NxN matrices. It gives,
\begin{equation}
    K_{T} = A+B+C+D
\end{equation}

\begin{equation}
   A =\begin{bmatrix}
        a&0&\dots&0\\
        \vdots&&\ddots&\vdots\\
        0&0&\dots&0\\
    \end{bmatrix}_{NxN}
    B =\begin{bmatrix}
        0&\alpha_1&\dots&d\alpha_N\\
        d\alpha_2&0&\dots&0\\
        \vdots&\vdots&\ddots&\vdots\\
        d\alpha_N&0&\dots&0\\
    \end{bmatrix}_{NxN}
\end{equation}

\begin{equation}
   C =\begin{bmatrix}
        0&\dots&\dots&\dots&0\\
        \vdots&\alpha^2_2&\alpha_2\alpha_3&\dots&\alpha_2\alpha_N\\
        \vdots&\vdots&&\ddots&\vdots\\
        0&\alpha_N\alpha_3&\dots&\dots&\alpha^2_N
    \end{bmatrix}_{NxN}
   D =\begin{bmatrix}
        0&\dots&\dots&\dots&0\\
        \vdots&\beta^2_2&\beta_2\beta_3&\dots&\beta_2\beta_N\\
        \vdots&\vdots&&\ddots&\vdots\\
        0&\beta_N\beta_3&\dots&\dots&\beta^2_N
    \end{bmatrix}_{NxN}
\end{equation}
which yields the following volume of circumstance for $K_{T}$:

\begin{equation}
    K_{T} \to 1+(N-1)(N+1)\text{ free coefficients}
\end{equation}

Introducing the mass term
\begin{equation}
   M_{T} \equiv \begin{bmatrix}
        0&\dots&\dots&\dots&0\\
        \vdots&m_{22}^{2}&m_{23}^{2}&\dots&m_{2N}^{2}\\
        \vdots&\vdots&&\ddots&\vdots\\
        0&m_{N2}^{2}&\dots&\dots&m_{NN}^{2}
    \end{bmatrix}_{NxN}
\end{equation}
one gets the following volume of opportunites for quanta masses at spin-1 sector 
\begin{equation}
    K_{T} +M^{2}_{T} \to 1+(N-1)(\frac{3}{2}N +1)\text{free coefficients}
\end{equation}

Considering the spin-0 sector, the $K_{L}$ matrix expressed at (C.9) is rewritten as

\begin{equation}
    K_{L} = E + F + G
\end{equation}

\begin{equation}
   E =\begin{bmatrix}
        0&\dots&\dots&\dots&0\\
        \vdots&8\beta^2_2&8\beta_2\beta_3&\dots&\beta_2\beta_N\\
        \vdots&\vdots&&\ddots&\vdots\\
        0&8\beta_N\beta_2&\dots&\dots&\beta^2_N
    \end{bmatrix}_{NxN}
   F =\begin{bmatrix}
        0&\dots&\dots&\dots&0\\
        \vdots&\beta_2\rho_{2}&\beta_2\rho_3&\dots&\beta_2\rho_N\\
        \vdots&\vdots&&\ddots&\vdots\\
        0&\beta_N\rho_2&\dots&\dots&\beta_N\rho_{N}
    \end{bmatrix}_{NxN}
\end{equation}
and
\begin{equation}
   G =\begin{bmatrix}
        0&\dots&\dots&\dots&0\\
        \vdots&8\rho^2_2&8\rho_2\rho_3&\dots&\rho_2\rho_N\\
        \vdots&\vdots&&\ddots&\vdots\\
        0&8\rho_N\rho_2&\dots&\dots&\rho^2_N
    \end{bmatrix}_{NxN}
\end{equation}
which gives,
\begin{eqnarray}
    K_{L}\to (N-1)(2N-1) \text{ free coefficients}
\end{eqnarray}
Adding the longitudinal mass matrix, one gets
\begin{equation}
    K_{L} +M_{L}^{2} \to (N-1)(\frac{5}{2}-1) \text{ free coefficients}
\end{equation}

Expressions (4.7) and (4.12) are showing the chances which the constructivist abelian model derived from eq. (2.6) provides for accommodating the quanta diversities.

Chance is a property under directive and circunstance. The directive is given by the minimum action principle and gauge symmetry determinations as Noether theorem, tensor energy-momentum invariance, gauge fixing, Ward identity. The circumstance by the free coefficients attached to the Lagrangian gauge scalars. As biology provides accidents and quantum mechanics probability, a free will is expressed by the constructivist volume of circumstances.

Consider the mass case. Given that the physical mass is determined perturbativily as the pole of a two point Green's function, the corresponding fields masses quanta are expressed through directive and circumstance. Eq. (4.1) contains as directive the presence of a massless quantum and as circumstance the existence of massive quanta depending on the theory free parameters. It is mandatory a massless gauge boson, $\mathbf{m}_{11} = 0$, due to the directive gauge symmetry. Circumstantial masses are determined depending on the theory free coefficients paraments as $m_{ij}^{2}$, $a_{I}$, $\mathbf{e}_{I J}$ and so on. Physical masses with chance of taking different values without violating gauge invariance.

\section{Evolution physics}
${}$\indent

Nature is made by transformations. However, more than an atomism devoted ultimate constituents and respective interactions, its objective is to individualize objects. Physics has to understand this determination. How to give character to constituents. A requirement asking for what is the dynamics able to do that.

A physics is being constructed in terms of set action. An associative physics where the variable is the variety. A perspective where physics converges to biology. While biology produces varieties (species) through genes and environment, physics generates (particles, etc) from last constituents and fields set. Both converging to the individuation proposed under environment. 

Physics is not only expressed through isolated dynamics but also under a set determinism. The fields set approach is expliciting an environmental physics. Showing that $\{G_{\mu I}\}$ interrelationships develops granular and collective fields strengths under an antireductionist dynamics. An associativity physics prospected by systemic equations. 

Quantum evolution is introduced by quantum systems evolving through set determinism. Quantum constituted through fields set environment. And so, the whole quantum associated to a given field $G_{\mu I}$ inserted in the set $\{G_{\mu I}\}$ is modifying according to the constructivist Lagrangian L(G) and correspondent classical equations studied at section 10. 

A quantum system is generated under symmetry of difference. An environmental bioogical physics is proposed. A whole quantum is individualized over constructivist lagrangian. A set determinism is established introducing varieties with diversity, growth, and chance. The meanings of evolution is introduced. A new picture to be understood from set action.

Eq. (2.6) provides physics with environment. Quantum systems with individualized quanta are formed. Differentiated energy packets are constituted. Acording to biology the corresponding diversity, growth and chance propitiate evolution. The physics of evolution is produced by varieties in quantum systems. Emergence follows as evolution under circumstance. Complexity as the final product of the evolution manifestation.

\subsection{Quantic evolution}
${}$\indent

Particle physics has been collecting elementary particles. However, nature is more than particles and interactions. There is a physics beyond the excelse standard model particles protagonism. A grouping physics which associativities produce realistic varieties. Elementarities responsible for matter behaviour. Performing evolution.

At usual physics, a field is a function of space-time with an equation of motion which describes how this function varies. A particle is defined as its smallest vibration (quantum). So relativistic fields are the fundamentals elements of reality. However, at present scope they are just retreating on the particles dynamics and not their transformation.

Quantic evolution should be identified as particles transformations. They are verified at many particles reactions. Physics provides three perspectives for particle evolution qualified through quantum numbers modifications. First, physical entities changing under the renormalization group equation [34]. Second, fields modifications from constitutive fields equations coming from the fields set $\{G_{\mu I}\}$ interdependence [35]. Third, from a constructivist lagrangian, eq. (3.25) working as a quantum tree.

Thus, there is still an evolutive timeline to be understood. Understand field theory more than a fixed quantum moving in space-time. Introduce a dynamics mutating the quantum numbers values. A first perspective is with the renormalization group equation. To define a theory we most specificy the energy scale, say $\mu$, under the condition that physics can not depended on $\mu$. It yields the following equation. 
\begin{eqnarray}
    \frac{d}{d\mu}P(g, \mu) = (\mu \frac{\partial}{\partial \mu}+ \beta(g)\frac{\partial}{\partial \beta})P(g, \mu) 
\end{eqnarray}
where $P(g, \mu)$ is a physical funcation and $\beta$ the beta function given by $\beta(g) = \mu \frac{\partial}{\partial \mu}g$.

It develops parameters flowing  with energy scale $m(\mu)$ and e$(\mu)$. Showing how the fine structure constant $\alpha = \frac{1}{137}$ suffer modifications. 

By second, physical changes are also configurated through the constructivist equations of motion. At eq (9.1) a given field $G_{\mu I}$ is transforming under the fields set interdependence. Fields interdependence modifies the original reductionist field equation. For instance, consider the Maxwell photon.  

Thus, there is the set action principle of evolution to be studied. It incorporates two features beyond the renormalization group. The fields set $\{G_{\mu I}\}$ modified constitutive equations and the quantum numbers modifications stipulated by L(G) according to previous section. A third evolutive possibility is to consider the constructivist Lagrangian $L(G)$. The evolutive feature is a an inner dynamics in L(G) transforming their quantum numbers. It develops quantum numbes under environment, diversity, growth, chance. The conditions to provide individuations according to biology. 
 
A constitutive photon equation was obtained [35]. A new photon equation is obtained rewriting eq. (10.1) in terms of masses and currents. It yields the following photon field. It modified equation for an electromagnetic quantum system involving the electric charge intermediated by four intermediated bosons.

\begin{eqnarray}\label{equação de movimento A II}
 && \partial_{\nu}\left\{4a_{1}F^{\nu \mu} + a_{2}\left( \mathbf{e}^{[12][\nu \mu]} + \mathbf{e}^{[+-][\nu \mu]}\right)\right\}+ l^{\mu}_{AT} + c^{\mu}_{AT}=\nonumber
  \\
  &&= 2a_{3}\mathbf{m}^{2}_{U}U^{\mu} + j^{\mu}_{AT} +j^{\mu}_{UT}+ J^{\mu}_{NT}
 \end{eqnarray}

The above equation does not consider on isolated photon but inserted in a whole. It retreats the photon in an electromagnetic set. But, preserving the original QED symmetry. The coupling constant $\alpha$ between the electron and photon is preserved.
 
A mass type term appears. It is through the London term, $l^{\mu}_{AT}$. It constituted by fields association expressed as

 \begin{eqnarray}
      &&l^{\mu}_{AT} = \{4\mathbf{e}_{[12]}U_{\nu}U^{\nu}-2a_{3}(q_{2}^{2} +q_{1}q_{2})V_{\nu}^{+}V^{\nu -}\}A^{\mu}+ \nonumber
      \\
      &&\{4a_{3}\mathbf{e}_{[12]}A_{\nu}A^{\nu}-2(q_{1}q_{2} +q^{2}_{1})V_{\nu}^{+}V^{\nu -}\}U^{\mu}
    \label{london Amu}
 \end{eqnarray}
with $A_{\nu I}A^{\nu I}=\phi_{I}$, introducing means a scalar field with $mass^{2}$ dimension.

Term $c^{\mu}_{AT}$ means the conglomerate $mass$. It involves the conglomerates generated by fields
 \begin{eqnarray}
      &&c^{\mu}_{AT} = 4\mathbf{e}_{[12]}[A_{\nu}U^{\nu}U^{\mu} + a_{3}A_{\nu}U^{\nu}A^{\mu}+ 1/2(1+a_{3})(\mathbf{e}^{[12-][\mu \nu]}V_{\nu}^{+}  \nonumber
      \\
     && + \mathbf{e}^{[12+][\mu \nu]}V_{\nu}^{-})]-2q_{1}q_{2}( +V_{\nu}^{+}U^{\nu}V^{\mu-}+a_{3}V_{\nu}^{+}A^{\nu}V^{\mu-} +V_{\nu}^{+}U^{\nu}V^{\mu -} \nonumber
      \\
      &&+a_{3}V_{\nu}^{+}A^{\nu}V^{\mu -})+2\mathbf{e}_{[12]}(1+a_{3})(\mathbf{e}^{[12-][\mu \nu]}V_{\nu}^{+} + \mathbf{e}^{[12+][\mu \nu]}V_{\nu}^{-}) \nonumber
     \\
      &&-q_{1}^{2}V_{\nu}^{+}A^{\nu}V^{\mu -} -q_{1}^{2}V^{-}_{\nu}A^{\nu}V^{\mu +}\nonumber
      \\
      &&-2a_{3}q_{2}^{2}(V_{\nu}^{+}U^{\nu}V^{\mu -} +V^{-}_{\nu}U^{\nu}V^{\mu +})
 \end{eqnarray}
with $A_{\nu I}A^{\nu}_{J} = c_{IJ}$, introducing another kind of scalar field with $mass^2$ dimension. Notice that, eqs (12.2 - 12.3) are like the Higgs. They are introducing a mass perception through scalar fields.

The influence of these effective masses, $l_{\mu}$ and $c_{\mu}$, on the photon should be studied by the dispersion relation corresponding to eq. (12.1). Although the classically the photon has mass zero and quantically protected by the Ward identify, there is for such abelian fields set a contribution to be understood through the correspondent dispersion relation [36].

The current term is expressed through the interaction between fields strengths and potential fields. It gives,
 \begin{eqnarray}
      &&j^{\mu}_{AT} = \mathbf{e}_{[12]}[(a_{1}F^{\mu \nu} + u_{2}U^{\mu \nu})(U_{\nu} + a_{3}A_{\nu})+ 1/2\mathbf{e}_{[12]}(1+a_{3})(V^{\mu \nu +}V_{\nu}^{-}+ V^{\mu \nu -}V_{\nu}^{+})]\nonumber
      \\
      &&-(2iq_{1}+a_{3}q_{2})(\partial^{\mu}V_{\nu}^{+}V^{\nu -}-\partial_{\nu}V_{\mu+}V^{\nu -} +\partial^{\mu}V_{\nu}^{-}V^{\nu +} - \partial_{\nu}V^{\mu-}V^{\nu +} )\label{corrente constitutiva A}
 \end{eqnarray}
and
 \begin{eqnarray}
      &&j^{\mu}_{UT} = \mathbf{e}_{[12]}[(a_{1}F^{\mu \nu} + u_{2}U^{\mu \nu})(A_{\nu} + u_{3}U_{\nu})+ 1/2\mathbf{e}_{[12]}(1+u_{3})(V^{\mu \nu +}V_{\nu}^{-}+ V^{\mu \nu -}V_{\nu}^{+})]\nonumber
      \\
      &&-(2iq_{1}+u_{3}q_{2})(\partial^{\mu}V_{\nu}^{+}V^{\nu -}-\partial_{\nu}V_{\mu+}V^{\nu -} +\partial^{\mu}V_{\nu}^{-}V^{\nu +} - \partial_{\nu}V^{\mu-}V^{\nu +} ).
 \end{eqnarray}

Eq. (12.4-12.5) are showing an electromagnetism with potential fields physicalities. A nonlinearity with potential fields interacting with granular and collective fields. A physics expected by Faraday, Bohm-Aharanov and Aharanov-Casher effects [37].

Noether current is related as

\begin{eqnarray}
 J^{\mu}_{NT} \equiv iq\{V_{\nu}^{+}[v_{1}V^{\mu \nu -} + v_{2}\mathbf{e}^{[12-][\mu \nu]}] -V_{\nu}^{-}[v_{1}V^{\mu \nu +} +v_{2}\mathbf{e}^{[12+][\mu \nu]}]\}\nonumber
\\\label{Noether corrente Eq. 5.10}
\end{eqnarray}
where $q$ means the electric charge. Notice also the presence of uncharged fields coupled to electric charge.

Thus, eq. (12.1) is expressing the photon constructivist transformation. A photonics evolutive abelian fields dynamics. A whole quantum photon evolving at space-time. Although a massless photon is preserved by gauge invariance, constructivist effects are added on its behaviour. New terms on mass and charge are introduced through classical interdependent equations. The electron-photon interaction is preserved, but, the photon behaviour is added by introducing an induced mass through London and conglomerate terms and with interactions beyond electric charge and potential fields. There is a mass term eq (12.2) to be verified by the respective photon dispersion relation. Perhap a darwinian type quantum evolution

A quantum with potential to become something else is proposed. A mutating physics proposed under set action. While at reductionist methodology a thing is what it is, and the corresponding equations of motion just act on their space-time dynamics, the constructivist lagrangian introduces a new performance. The whole quantum is expressed in terms of diversity, nonlinearity, growth interdependence, chance, directive and circumstances. A similarity to Darwin evolution.

Thus, a given field $G_{\mu I}$ dynamics is not to be interpreted just as a wave equation. It contains a quantum numbers environment which is changing under a set action space-time dynamics. From biology, evolution happens through cambial exchange between generations. Here, the part-whole duality generates a conglomerates lineage. Instead of generations, it yields successive sets  through. The biology conditions to propiate evolution - diversity, growth, chance - appears in physics. Modifying the particle structures by as biology do with species. And so, physical entities as masses, charges, coupling contant will be changing according to the fields environment context.

Darwin shows that evolution arises from environment. It yields ingredients as variety, diversity, interdependence, self-interaction, growth chance, adaptation. The elements of the darwinian type of evolution that make up natural selection. A comparision to be investigated. Extra polet whether the darwinian natural solecation is the nature standard way to move from single to complex. Mutatis mtuandis, the challenge for physics should be to find quantum evolution within darwinian standards.

Considering that biology and physics are natural sciences, there must be a correlation between the formation and determination of species and particles. The difference is that in his two books Darwin did not use mathematics to show the evolution of species [16]. Meanwhile, physics has to structure these concepts analytically. Contemplate evolution not by figures, but mathematically.

In the symmetry of difference physics finds its mathematical prescription for describing evolution. Based on the antireductionist gauge theory, it elaborates the physicality of the elements that lead to Darwin's natural selection. It develops an environment through the set of fields $\{G_{\mu I}\}$, differentiated quanta produced by L(G), evolutionary determinism expressed by respective differential equations.

In this way, the symmetry of difference is a candidate to develop the notion of evolution in physics. Sections 8-11 mathematize evolutionary precepts. A systemic dynamics in space-time is provoking the sense of transformation of the field itself. This would be the perspective of a constructivist lagrangian. Constituting changes in its quantum numbers. Promoting a new order, where, instead of the space-time dynamics of particles, introduce the physics of particles evolution.

The symmetry of difference not only brings togetherness and diversity. It provides an arrow of time with the sense of evolution. Introducing another feasible modification of quantum physics. The correspondent fields equations express joint action, interdependence between fields, growth through non-linearity, chance under free coefficients. Adaptation is driven by boundary conditions and energy minimization. 

An evolutive quantum system is performed. The fields set $\{G_{\mu I}\}$ generates a whole quantum evolving their quantum numbers through a space-time dynamics. Producing a physics with particles variations and corresponding palaeontology. Since 1930s physics knows that particles are not only interacting and moving at space-time. While in the past physics moved from mechanics to field theory, now we have to introduce an associative field theory mutate under a quantum tree.

\subsection{Quantum palaeontology}
${}$\indent
A particle contains an history to be told. It is not a fixed object moving through space-time. Particles have an origin and a process. Since Big Bang have been forming diverse structures. And so, the particles quantum numbers transformations is a physics to be interrogated.

Nature is the science of changing. The second Newton's law narrates the space-time dynamics of a body. However, there is another aspect of modification which is the inner transformation. When this body is replaced by a nature internal transformation. A quantum mutation to be understood by field theory. Physics has to investigate how a certain quantum transforms its structure. 

The LHC is projected to discover particles. Its essence is the particles production. However, evolutive laws has to be considered. Look at LHC experiments not only as the search for new particles, but also, regarding how particles are developed along time. Search for the tracks left by their mutations. Particle physics contains an evolutive perspective to be studied. The quantum is not an entity repeating the same interactions. It contains an inner dynamics. Understand how the formative quantum numbers are modified. A change to be registired by a quantum numbers paleontology.

A register for quantic evolution is expected. Identify the evolution tracks left by eq. (3.29). Antireductionist properties were qualified at previous sections. A study was expressed about the whole quantum transformations. The difference between a field treated by an isolated differential equation and inserted in an interdependent fields system. Undersand the evolutive environment expressed by a coupled differential equations system.

A quantum system is being proposed through constructivist lagrangian. The symmetry of difference creates an environment with quanta diversity working as a set. The associated fields develops a self-organizing system. An inner whole space-time dynamics is constituted. Particles mutations are derived. Properties as diversity, interdependence, nonlinearity, chance, directive and circumstance are promoting internal transformations. A new physics is provoked. It is to register on the particle structure modifications.

We should explore on quantum paleontology. Consider the conglomerate physics. Following the dynamics expressed from a constructivist abelian lagrangian, it provides set transformations features. Fields strengths at eq. (9.1) are acting as sources for modifying their own values. A selfgrowth set is produced under potentiality and nonlinearity expliciting a development to be analysed by inner quantum numbers modifications.

Physics has to identify on elementary particles palaeontology. Quantum numbers are transformed by a set action and lineage. An evolutive quantum system is formed instead of an amorphous concept of the electron, an appendage of elementarity, physics makes the electron change from ultimate constituent to variety. Associating the single electron with quantum systems as leptonic families, Cooper pairs and so forth. A new electronic expression in nature depending on set action.

Nature evolution can be detected microscopically. There is a quantum  physics to be registered by palaeontology. The whole quantum variety dynamics under potentialities and realities produces quantum transformations. A quantum historical process is defined. Eq. (26) involves set properties composing the quantum evolution. The correspondent quanta diversiy introduces an inner space-time transformations. Quantum numbers changing are leaving palaeontological tracks to be investigated.

\subsection{Complexity}
${}$\indent

Experimental physics detected the universe between $10^{-18}m$ (quarks) until $10^{26}m$ (distant galaxies). A long road containning a process between fundamental physics and complexity. Since an atom or a cell the universe produce a linage of conglomerates. A process achieving solar systems and species. A spectrum challenging physics to understand how the universe moves from simplicity to complexity.

Physics and biology are natural sciences. And so, whole quantum should share the meaning of variety in biology. A clue for, while the theory of evolution in biology is based on environment and varieties, the physical world is in terms of fields set and whole quanta. As a cell, the whole quantum contains a whole unity. Parts in the whole are working with diversity and interdependence. Making the atomist quantum move from force and interaction to quantum systems and interrelationships. Producing multiplication under nonlinearity and opportunities. And, introducing complexity as a next theme from conglomerate physics.

The complexity scenario emerges. Physics has to explain how nature evolves from a single particle to complex systems. There is a complex get together developed by constructivist lagrangian. Quanta under antireductionist coalization. Parts in the whole are candidates to express physics under complexity. Expressing functionalities under antireductionist laws.

Thus, complexity leads physics to new properties than fundamental physics. Features which will allow assemble diverse quanta in a physical structure beyond fundamental law. Further consequence from evolution, quantum complexity context is the next topic to be developed by quantum field theory. Introduce a quantum under the synapsis presence.

\section{Quantum system from gauge symmetry}
${}$\indent
An associative quantum field theory is proposed. It leads physics beyond the usual atomist quantum field theory initiated in 1927 by  Dirac [38]. Introduces through gauge symmetry the passage from a single quantum to a quantum system. Discover a continuity between reductionism and antireductionism. 

A third type of quantum is expressed. It is called whole quantum. An excitation derived from a field set. An energy signad made from a quantum system. While fundamental physics is based on ultimate constituents a quantum system develops varieties as elementaries. As biology, from proton onwards,  physics would be moved by varieties inserted in conglomerates.

\begin{figure}[H]
    \centering
\begin{tikzpicture}
        \node at (0,4.3) {single quantum};
        \node at (6,4.3) {quantum system};
        \draw[->] (1.5,4.3) -- node[above] {gauge} (4.5,4.3);
        \node at (3,4) {symmetry};
    \end{tikzpicture}
    \caption{Constructivist Lagrangian makes the passage from a single quantum to a quantum system.}
\end{figure}
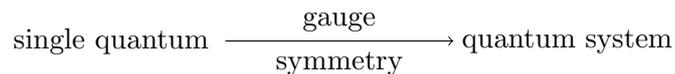

The P.W. Anderson lemma 'more is different' is still to be understood [39]. Quantum systems are well identified structures in nature by constructivist lagrangian. Enlarging field theory from Dirac to Anderson. Consider a quantum of associated quantum system with whole properties. 

Diverse realistic situations provide quantum systems. For electron, one gets the following cases: superconductivity with Cooper pairs [40], electronic collective excitations separating the electron charge and spin and introducing the so-called holon and spinon [41], plasmons in plasma physics [42], polarons [43], polariton [44], excitons [45]. For spin, waves as phonons [40]. For photon, quantum system in crystals [47]. For electromagnetism, four bosons  with an electromagnetism based on four photons associations (massless, massive, charged) [33]. For EM phenomena, Hall effect [48].

Thus, physics should  study quantum system through gauge symmetry and the challenge is how to move from a single to a system of particles. The discussion is not on their existence but about the approach. Current physics states that a particle in a quantum system instead of classical properties exibits a behaviour ruled by quantum mechanics laws. Described by wave functions under the prescription of possibility measurement.

The fields set action introduces a new interpretation for quantum system. Field theory appeared as an extension to quantum mechanics but was not able to describe quantum mechanics but was not able describe quantum systems. The associated quantum field theory appears to cover this gap. The pluriformity of quantum diversities analysed at this work support a quantum system formulation without requiring quantum mechanics. Diversely from wave function its description is in terms of variables coming from mechanics and waves. Based on symmetry of difference a quantum system is proposed where conglomerates are described by measurable variables.

This is the physics promoted by eq. (2.6). The correspondent antireductionist gauge symmetry relates the single and the set as required. A way where the particle is inserted in the whole, preserving its atomist symmetry, and with new collective properties. Also, maintaining the Planck quantum variables. The a quantum system as a single quantum with antireductionist physicalities.

Four features make such passage from an isolated particle to a system particles. First, the enlargement from atomism to constructivism preserving the original gauge symmetry. Second, a quantum diversity without increasing the group theory rank, or, requiring spontaneous symmetry breaking. Third, a collective physical description in terms of the physical variables that constitute the du Broglie quantum. Finally, a whole quantum with constructivist properties unrealized by quantum mechanics.

Thus, an evolutive quantum system is proposed. As limit, their features preserve the atomist quantum symmetry, and, includes antieductionist properties. Proposing a condensed matter physics without requering quantum mechanics. Considering the whole as fundamental unity, it introduces a set physics under diversity, interdependence nonlinearity, chance.  A quantum system with space-time transformation involving growth, evolution, emergence, complexity.

\section{Conclusion}

${}$\indent

At last 2500 years physics has been searching for the greek atom. Understand nature under ultimate constituents. Over this perspective, quantum physics became the study of the energy packet. Qualifying the minimum energy structure under Planck constant. However, nature performed by quantum mechanics is unsatisfactory. It does not reach its full maturity. It appears as framework for subatomic particles, but contains various not understandable fuzzy aspects. The quantum world still requires a further explanation.

The quantum realm remains a debate. Nature features as associativity, confinement and complexity lead physics to a set order. Gauge symmetry assembling fields $\{G_{\mu I}\}$ is a new procedure. Constructivist Lagrangian as the framework for set action. The quantum numbers that constitute a quantum no more specified by reductionist properties. Determined in terms of the involved whole.

Thus, the quantum system may be interpreted beyond quantum mechanics. A view pointing out to a third moment on quantum story. As relativity extends classical mechanics, the current reductionism and quantum mechanics should be transposed to constructivism. The last constituents not only form particles, but make quantum systems as nuclei, atoms, molecules. A perspective for a third quantum be introduced by an associative physics enlarging interactions to interrelationships.

Three types of quantum are classified. Granular, quantic and whole. The first two have historic signature. However, nature from its capacity to generate systems and patterns, expresses a further energy packet. The whole quantum as a next possibility. Distinct arguments may select its existence. Three thesis we select to support a third quantum physics. Aristotle, quantum mechanics, variety. They produce arguments from where the quantum should not be restrict to Planck and wave-particle.

Aristotle divided three modes of being. Really being, not-being, potentially-being. Given such generic statement, let us consider its application on quantum physics. Understand on the individualization possibilities. In terms of quantum, the first one was detected experimentally and expressed theoretically  by the atomist quantum field theory. Quarks under confinement introduce the not-being, a quantum detected indirectly. The potentially-being is left for constructivism.

On that, there are two methodologies for understanding the quantum nature. Atomist and constructivist. Atomism prevaleces at current physics. Ultimate constituents, as quarks and leptons, working like matryoshkas make physics as the summatory of parts. Atomism describes the first two Aristotles assumptions. On the other hand, parts associated in the whole introduce another type of energy packing. Therefore, for answering the third Aristotle philosophical inquiring, one deduces that, there is a quantum constructivism to be investigated.

More is different, and so, the constructivist physics should be taken. Consider on the quantum associativity. An extension where the $A_{\mu I}$ field will work inserted at fields set $\{A_{\mu I}\}$. A passage from reductionism to antireductionism preserving the original symmetry. An enlargement with interactions preserving the corresponding covariant derivative, but, establishing new antireductionists properties for the field $A_{\mu I}$. For instance, at QED the electron-photon interaction is preserved by $(\partial_{\mu}+ieA_{\mu})\psi$, but the photon properties differentiated from usual electromagnetism as section 5 studied.

A second indication on a third quantum physics would be due to quantum mechanics uncompleteness. A precious debate happened in 1935 with the Einstein, Podolski, Rosen paradox. The paper title was 'Can Quantum-Mechanical Description of Physical Reality Be Considered Complete?' [49]. The answer was no. In back, Bohr said yes [50]. A discussion followed, Schrödinger introduced the cat [51]. Later on, other contribuitions came from Bohm [52], Bell [53] and others.  

Physics is waiting for an undestanding of the laws of nature beyond wave function. Quantum mechanics effects as probability [54], tunneling [55], entanglement [56], coherence [57] are still expecting a better comprehension. Instead of describing the quantum phenomenon through the mathematical entity psi, which does not have a direct physics, consider explanations through measurable variables. The quantum description under a new perspective. Particles quantum numbers as spin, mass, charges, discrete symmetries expressed over acessible variables than quantum mechanics operators methodology. 

Quantum systems should be performed under the symmetry of difference acessible physical entities. The quantum particles behaviour not necessarily described by Schrödinger equation. Replace the abstract wave psi in phase space by the whole quantum. The physical quantum picture of reality expressed by dynamical readable variables. Something is expected. Quantum effects written in terms of du Broglie entities as energy-momentum, wavelength. For instance, entanglement, where two particles can be linked together even separated at large distance, treated with a dispersion relation depending on palpaple entities [36].

The third support for a next quantum physics is the variety physics. A whole physics introducing a second type of individuation. The whole elementarity called as varieties. The associative principle makes the ultimate constituents express a constructivist behaviour. Moving physics from excelse ultimate constituents to ordinary varieties description. Introducing varieties with antireductionist properties. A qualitative change happens with respect to classical physical and quantum mechanics. A variety physics with new quantum features.

A quantum system is defined through varieties. The crucial difference for the atomist quantum physics is a set action producing. Five aspects not expressed by the atomist quantum are constructing a quantum system through gauge symmetry. It includes five features. They are set action, antireductionist properties, nonvirtuality, functionality, evolution.

Thus, a new definition on quantum system is given by constructivist physics. Firstly, the set action introduces the meaning of associativity. The ultimate constituents not only form particles as quantum systems based on varieties. Secondly, such quanta webfication introduces antireductionist laws. Relationships as interaction, induction, connection are enlarging reality through the set action. Deriving a whole quantum under a set determinism. Providing set, diversity, interdependence, nonlinearity, chance. Propitiating growth, evolution, emergence, complexity.

A third feature is nonvirtuality. Physics is being based on virtual particles. There is a not very clean mixing effects between virtual and real in quantum field theory. Although the lagrangian is real, the related vaccuum and the corresponding Feynman graphs are performed under virtualities. Examples are photon-photon scattering through QED and Higgs decay through Standard Model. 

Thus, a constructivist lagrangian is augmenting reality. It provides a more accurated realism than atomist lagrangians. A pertubative quantum reality is propiated through gauge invariant terms. Realistic Feynman graphs are performed. Working on it, with pieces of reality which are nonvirtual gauge invariant terms, it develops a real pertubative theory. Events described by Feynman graphs based at reality.

Fourth, we would observe that a quantum embedded in a whole provides quantum functionalities. Physical entities as spin, mass, charges and so forth, are not in service of four interactions. They are working as funcionalities. Depending on how the quantum is associated in the whole. A physics where quantum functionalities replaces four interactions to a constructive scope.

Finally, the constructive physics provides nature with its most relevant behaviour, which is evolution. Evolution is not an usual concept at current physics. Equations are just describing entities moving at space-time. They did not consider on how particles are modified internally. A difference appears through grouping physics. The fields set action develops the quantum mutation. It provides a set dynamics transforming their elementarities (varieties) internal struture. Evolving the correspondent quantum numbers. A physics where ultimate constituents are preserved and varieties modified.

The constructivist physics provides the ingredients for quantum evolution. They are environment (set action), set arrow (directive), interdependence (interwined potential fields), growth (potentiality, nonlinearity), chance (free will, volume of circumstances), adaptation (minimal energy, boundary condition). The antireducticionist gauge symmetry determines an environmental physics providing the quantum internal change.

A quantum evolution happens. Eq. (2.6) introduces the meaning of evolution. A physics further than usual determinism. It flavors the quantum take decisions on its formation inside of a fields set environment. Fields equations are more than moving at space-time they carry quantum numbers. Evolution objective is the matter transformation. Give choice for objects existence. Manifest individuation. A physics establishing quantum mutation and correspondent quantum numbers paleontogy. It provides an individuation dynamics where each corresponding quantum evolves through an individuation. 

Concluding, a third quantum physics is expressed by an associative quantum field theory. It introduces a physics beyond the usual atomist quantum field theory. Its objective is to replace quantum mechanics. A perpective is expected for covering quantum mechanics properties without the psi wave function. Introduce quantum evolution through gauge symmetry. Instead of wave function it introduces the antireductionist symmetry determines. It the symmetry of difference. The whole-part complementarity becomes a quantum physics to be studied. The whole quantum is qualified with antireductionist properties without requering quantum mechanics.

\begin{appendix}

\section{Apendice A: constructive basis}

${}$\indent
Abelian gauge symmetry contains just one gauge parameter. Based on it, a fields family is associated under the gauge transformation

\begin{eqnarray}\label{Tranformation bases}
    A^{'}_{\mu I} = A_{\mu I} + k_{I}\partial_{\mu}\alpha 
\end{eqnarray}

Considering that the achievenment is to construct a gauge invariant theory, one redefine eq. (\ref{Tranformation bases}) in terms of the constructor basis $\{D, X_{i}\}$

\begin{eqnarray}
&&D_{\mu}=\sum_{I =1}^{N} A_{\mu I}\nonumber\\
&&X_{\mu i}=A_{\mu\,i+1}-A_{\mu\,i}; i=2...N
\end{eqnarray}
which yields the following gauge transformations
\begin{eqnarray}
&&D_{\mu} \longrightarrow D'_{\mu}=D_{\mu}+N\partial_\mu{\alpha}\nonumber\\
&&X_{\mu i} \longrightarrow X'_{\mu i}=X_{\mu i}
\end{eqnarray}
generating the following Lagrangian
\begin{equation}\label{Lagrangian constructor basis}
    \lag = Z_{[\mu \nu]}Z^{[\mu \nu]}+Z_{(\mu \nu)}Z^{(\mu \nu)}+Z_{[\mu \nu]}\Tilde{Z}^{[\mu \nu]}+\frac{1}{2}m^2_{ij}X_{\mu i}X^\mu_j
\end{equation}
where $\Tilde{Z}_{[\mu \nu]} = \epsilon_{\mu \nu \rho \sigma}Z^{\rho \sigma}$,  $Z^{[\mu \nu]}\Tilde{Z}_{[\mu \nu]}$ is called as the semitopological term.

\begin{eqnarray}
    Z_{[\mu \nu]} &=& d D_{\mu \nu} + \alpha_i X_{\mu \nu i}+ \gamma_{[i j]} X_{\mu i} X_{\nu j}
    \\
    Z_{(\mu \nu)}&=&\beta_{i} S_{\mu \nu i}+\rho_{i} g_{\mu \nu}S^{\alpha}_{\alpha i} +\gamma_{(i j)}X_{\mu i}X_{\nu j}+\tau_{(i j)}g_{\mu \nu} X_{\alpha i} X^\alpha_{\alpha j}
\end{eqnarray}
with the following gauge invariants fields strengths
\begin{eqnarray}
    D_{\mu \nu}&=&\partial_\mu D_\nu - \partial_\nu D_\mu \\
    X_{\mu \nu i}&=&\partial_\mu X_{\nu i} - \partial_\nu X_{\mu i}\\
    S_{\mu \nu i}&=&\partial_\mu X_{\nu i} + \partial_\nu X_{\mu i}\\
    S^\alpha_{\alpha i}&=&2\partial_\alpha X^{\alpha i}
\end{eqnarray}

Separating the above Lagrangian in pieces,
\begin{equation}
    \lag=\lag_K+\lag_I^3+\lag_I^4+\lag_{st}+\lag_{GF}+\lag_M \label{lagrangian 12}
\end{equation}
where each term contains antisymmetric and symmmetric sectors
\begin{eqnarray}
\lag_K &=& \lag_K^A +\lag_K^S\\
\lag_I^3 &=& \lag_A^3 +\lag_S^3\\
\lag_I^4 &=& \lag_A^4 +\lag_S^4\\
\lag_{st} &=& \lag_{st}^3 +\lag_{st}^4
\end{eqnarray}

Considering eq (\ref{lagrangian 12}), one gets the kinetic term

\begin{equation}
\lag_K^{A} = d^{2}D^{\mu\nu}D_{\mu\nu} + 2d\alpha_i D^{\mu\nu}X_{\mu \nu}^i+\alpha_i\alpha_j X^{\mu \nu}_i X_{j\mu \nu}
\end{equation}

\begin{equation}
\lag_K^{S}= 
\beta_i\beta_jS^{\mu \nu}_iS_{j \mu \nu }+2\beta_i\beta_jg_{\mu \nu}S^{\mu \nu}_iS_{j \alpha}^\alpha+4\rho_i\rho_jS^{\alpha}_{i \alpha}S_{j \beta}^\beta
\end{equation}

The trilinear interaction Lagrangian is

\begin{eqnarray}
    \lag^3_A &=& 2d\gamma_{[ij]}D^{\mu\nu} X_\mu^i X_\nu^j +2\alpha_i\gamma_{[kj]}X^{\mu \nu}_iX_{k\mu}X_{j\nu}\no\\
    &=& 4d\gamma_{[ij]}\partial^\mu D^\nu X^i_\mu X^j_\nu+4\alpha_i\gamma_{[kj]}\partial^\mu X_i^\nu X_{k\mu} X_{j \nu}
\end{eqnarray}

\begin{eqnarray}
    \lag^3_S &=& 4\beta_i\gamma_{(kj)}\partial^{\mu} X_i^\nu X_{k \nu}X_{j\nu} +4\beta_i\tau_{(kj)}\partial^\mu X_i^\nu X_{k\alpha}X^{i\alpha}+\no\\
    &+& 4\rho_i\gamma_{(kj)}g^{\mu \nu}\partial_\alpha X_i^\alpha X_{k \mu} X_{j\nu}+16\rho_i\tau_{(kj)}\partial_\alpha X_i^\alpha X_\beta^k X^{j \beta}
\end{eqnarray}

\begin{eqnarray}
    \lag^3_{st} = &-&4d^2\gamma_{[jk]}\epsilon^{\mu\nu\rho\sigma}\p_\mu X^j_\rho D_\nu X^k_\sigma+\no\\
    &-& 4d\alpha_i\gamma_{[jk]}\epsilon^{\mu\nu\rho\sigma}\p_\mu X^j_\sigma X^i_\nu X^k_\sigma+\no\\
    &+& 2\gamma_{[ij]}\alpha_k\epsilon^{\mu\nu\rho\sigma}\p_\rho X^j_\sigma X^i_\mu X^j_\sigma+\no\\
    &+& 2d\gamma_{[ij]}\epsilon^{\mu\nu\rho\sigma}\p_\rho D_\sigma X^i_\mu X^j_\sigma+\no\\
\end{eqnarray}

\section{Apendice B. Volume of circumstance}
${}$\indent
A new physicality coming from the fields set is the volume of circumstance. Given the above expressions we will build a table considering the number of free coefficients of each gauge invariant Lagrangian term. It yields the table below relating gauge scalars and free coefficients. It contains the volume of circumstance associated to each term.

For $\lag^{3}_{A}$:

\begin{table}[htbp]
  \centering
   \caption{Volume of circumstance $\lag^3_A$}
    \begin{tabular}{c|c|c}
    \hline
    Scalar & Coefficient & Nº of Coefficients \\
    \hline
    $\p^\mu D^\nu X^i_\mu X^j_\nu$ & $d\gamma_{[ij]}$ & $\frac{(N-1)(N-2)}{2}$\\
    \hline
    $\p^\mu X^\nu_i X^k_\mu X^j_\nu$ & $\alpha_i\gamma_{[kj]}$ & $\frac{(N-2)(N-1)^{2}}{2}$\\
    \hline
    \end{tabular}%
  \label{tab: L3a}
\end{table}

The volume of circumstance for $\lag^3_A$ is
\begin{equation}
    \frac{1}{2}(N-1)(N^{2}-2N)
\end{equation}

For $\lag^{3}_{S}$:

\begin{table}[htbp]
  \centering
   \caption{Volume of circumstance $\lag^3_S$}
    \begin{tabular}{c|c|c}
    \hline
    Scalar & Coefficient & Nº of Coefficients \\
    \hline
    $\p^\mu X^\nu_i X^k_\mu X^j_\nu$ & $\beta_i\gamma_{(kj)}$ & $\frac{N(N-1)^2}{2}$ \\
    \hline
    $g_{\mu\nu} \p^\mu X^\nu_i X^k_\alpha X^{j\alpha}$ & $\beta_i\tau_{(kj)}$ & $\frac{N(N-1)^2}{2}$ \\
    \hline
    $\p_\alpha X^\alpha_i X^k_\mu X^{j}_\nu$ & $\rho_i\gamma_{(kj)}$ & $\frac{N(N-1)^2}{2}$ \\
    \hline
    $\p_\alpha X^j_\beta X^{j}_\beta\mu X^{k\beta}$ & $\rho_i\tau_{(kj)}$ & $\frac{N(N-1)^2}{2}$ \\
    \hline
    \end{tabular}%
  \label{tab: L3s}
\end{table}%

The volume of circumstance for $\lag^3_S$:
\begin{equation}
    2N(N-1)^2
\end{equation}

The volume of circumstance for $\lag^3_{st}$:
\begin{equation}
    (N-1)(N^{2}-2N)
\end{equation}

Number Total for $\lag^{3}_{T}$:
\begin{equation}
    \frac{1}{2}[N(5N-8)+2](N-1)
\end{equation}

The quadrilinear Lagrangian is separated
\begin{equation}
    \lag_I^4 =\lag_A^4 +\lag_S^4 + \lag_{st}^{4}
\end{equation}
where
\begin{eqnarray}
    \lag_A^4 &=& \gamma_{[ij]} \gamma_{[kl]} X_\mu^i X_\nu^jX^{k\mu} X^{l\nu},
\end{eqnarray}

and
\begin{equation}
    \lag^4_{st} = \epsilon^{\mu\nu\rho\sigma}\gamma_{[ij]}\gamma_{[kl]}\p_\rho X^i_\mu X^j_\nu X^k_\rho X^l_\sigma
\end{equation}
Considering $\lag^4_A$,
\begin{table}[htbp]
  \centering
   \caption{Volume of circumstance $\lag^4_A$}
    \begin{tabular}{c|c|c}
    \hline
    Scalar & Coefficient & Nº of Coefficients \\
    \hline
    $X^i_\mu X^j_\nu X^{k\mu} X^{l\nu}$ & $\gamma_{[ij]}\gamma_{[kj]}$ & $\frac{N(N-1)}{2}$ \\
    \hline
    \end{tabular}%
  \label{tab: L4A}
\end{table}

The volume of circumstance for $\lag^{4}_{A}$:
\begin{eqnarray}
    \frac{N(N-1)}{2}
\end{eqnarray}

Considering $\lag^{4}_{S}$,

The volume of circumstance for $\lag^4_{S}$:
\begin{eqnarray}
    \frac{N(N-1)}{4}(N^{2}+2)
\end{eqnarray}

Considering $\lag^{4}_{st}$, the volume of circumstance for is
\begin{equation}
    \frac{N(N-1)}{2}
\end{equation}
    
\newpage

Finally,we have to establish the total volume of circumstance derived from the abelian fields expressed by the constructor basis $\{D_{\mu}, X_{\mu i}\}$

\begin{equation}
    \lag = \lag^A + \lag^S+\frac{1}{2}m^2_{ij}X^i_\mu X^\mu_j + \lag_{GF}
\end{equation}

\begin{table}[htbp]
  \centering
   \caption{Relationship between scalar gauge and No. of free coefficients}
    \begin{tabular}{c|c|c}
    \hline
    Scalar of Gauge & Coefficient & Nº of Coefficients livres\\
    \hline
    $D_{\mu\nu}D^{\mu\nu}$ & $d^2$ & 1\\
    \hline
    $D_{\mu\nu}X_i^{\mu\nu}$ & $d\alpha_i$ & (N-1)\\
    \hline
    $X^i_{\mu\nu}X_j^{\mu\nu}$ & $\alpha_i\alpha_j$ & $\frac{N(N-1)}{2}$\\
    \hline
    $S^i_{\mu\nu}S_j^{\mu\nu}$ & $\beta_i\beta_j$ & $\frac{N(N-1)}{2}$\\
    \hline
    $g_{\mu\nu}S_i^{\mu\nu}S^\alpha_{\alpha j}$ & $D_i\rho_j$ & $N(N-1)^2$\\
    \hline
    $S^\alpha_{\alpha i}S^\beta_{\beta j}$ & $\rho_i\rho_j$ & $\frac{N(N-1)}{2}$\\
    \hline
    $X^i_{\mu}X^\mu_{j}$ & $m^2_{i j}$ & $\frac{N(N-1)}{2}$\\
    \hline
    $\p^\mu D^\nu X_{\mu i}X_{\nu j}$ & $d\gamma_{[i j]}$ & $\frac{(N-1)(N-2)}{2}$\\
    \hline
    $\p^\mu X^\nu_i X_{\mu j}X_{\nu k}$ & $\alpha_i\gamma_{[i j]}$ & $\frac{(N-1)^2(N-2)}{2}$\\
    \hline
    $\p^\mu X^\nu_i X_{\mu k}X_{\nu j}$ & $\beta_i\gamma_{(i j)}$ & $\frac{N(N-1)^2}{2}$\\
    \hline
    $g_{\mu \nu}\p^\mu X^\nu_i X_{\alpha}^j X^\alpha_{k}$ & $\beta_i\tau_{(j k)}$ & $\frac{N(N-1)^2}{2}$\\
    \hline
    $g^{\mu \nu}\p^\alpha X^i_\alpha X_{\mu j} X_{\nu k}$ & $\rho_i\gamma_{(j k)}$ & $\frac{N(N-1)^2}{2}$\\
    \hline
    $\p^\alpha X^i_\alpha X_{\beta j} X^\beta_{k}$ & $\rho_i\tau_{(j k)}$ & $\frac{N(N-1)^2}{2}$\\
    \hline
    $X^\mu_i X^\nu_j X_\mu^k X_\nu^l $ & $\gamma_{[i j]}\gamma_{[k l]}$ & $\frac{N(N-1)}{2}$\\
    \hline
    $X^\mu_i X^\nu_j X_\mu^k X_\nu^l $ & $\gamma_{(i j)}\gamma_{(k l)}$ & $\frac{N(N-1)[N(N-1)+2]}{8}$\\
    \hline
    $g^{\mu\nu} X_\mu^i X_\nu^j X_\alpha^k X^{l \alpha}$ & $\gamma_{(i j)}\tau_{(k l)}$ & $\left[\frac{N(N-1)}{2}\right]^2$ \\
    \hline
    $X_\alpha^i X_j^\alpha X_\beta^k X^{l}_{\beta}$ & $\tau_{(k l)}\tau_{(k l)}$ & $\frac{N(N-1)[N(N-1)+2]}{8}$ \\
    \hline
    $\epsilon^{\mu \nu \rho \sigma} \p_\mu X_\rho^j D_\nu X^k_\sigma$ & $d^2\gamma_{[j k]}$ & $\frac{(N-1)(N-2)}{2}$ \\
    \hline
    $\epsilon^{\mu \nu \rho \sigma} \p_\mu X_\rho^j X^i_\nu X^k_\sigma$ & $d\alpha_i\gamma_{[j k]}$ & $\frac{(N-1)^2(N-2)}{2}$ \\
    \hline
    $\epsilon^{\mu \nu \rho \sigma} \p_\rho D_\sigma X^i_\mu X^j_\nu$ & $d\gamma_{[i j]}$ & $\frac{(N-1)(N-2)}{2}$ \\
    \hline
    $\p_\mu D^\mu\p_\nu D^\nu$ & $\frac{1}{2\xi}$ & $1$ \\
    \hline
    $\p_\mu D^\mu\p_\nu X^{\nu i}$ & $\frac{1}{\xi}\sigma_i$ & $(N-1)$ \\
    \hline
    $\p_\mu X^{\mu i}\p_\nu X^{\nu j}$ & $\frac{1}{2\xi}\sigma_i\sigma_j$ & $\frac{N(N-1)}{2}$ \\
    \hline
    \end{tabular}%
  \label{tab:}
\end{table}%

The number of free coefficients of each term is:

\begin{eqnarray}
\lag_{A} &\to& 1 + \frac{1}{2}\left[N(N-1)+N+2\right](N-1)\\
\lag_{S} &\to& \frac{1}{2}\left[N^2(N-1)+4N(N\frac{1}{4})-2\right](N-1)\\
\end{eqnarray}

\begin{equation}
    \lag_{A+S} \to \frac{1}{2}\left[N^2(N+4)+(N-2)(N+1)\right](N-1)+1
\end{equation}
If we add the semitopological terms:
\begin{eqnarray}
\lag_{st}&\to& \frac{(N-1)}{2}\left[N^{2}-2\right]    
\end{eqnarray}

For mass term.
\begin{eqnarray}
\lag_{M} &\to& 1 + \frac{N(N-1)}{2}
\end{eqnarray}

Including the gauge fixing terms:

\begin{table}[htbp]
  \centering
   \caption{Volume of circumstance $\lag_{GF}$}
    \begin{tabular}{c|c|c}
    \hline
    Scalar & Coefficient & Nº of Coefficients \\
    \hline
    $\partial_{\mu} D^{\mu}\partial_{\nu} D^{\nu}$ & $\frac{1}{2\xi}$& $(N-1)$\\
    \hline
    $\partial_{\mu}X^{\mu i}\partial_{\nu}X^{\nu j}$&$\frac{1}{2\xi}\sigma_{i}\sigma_{j}$& $\frac{N(N-1)}{2}$\\
    \hline
    \end{tabular}%
  \label{tab: L4s}
\end{table}%
\begin{eqnarray}
    \lag_{GF} \to \frac{(N-1)}{2}(N+2)
\end{eqnarray}

Thus, the total number of free coefficients is
\begin{equation}
    \lag \to \frac{1}{2}\left[N^2(N+4)+(N-2)(N+1)+2\right](N-1)+2
\end{equation}

\section{Apendice C. Kinetic sector}
\indent${}$

Next topic at constructor basis is to undestand on the whole quanta diversity. Writing the kinetic lagrangian in terms of projection operators:

\begin{eqnarray}
    \lag_{K} &=& -2d^{2}D_{\mu} \Box \theta^{\mu\nu}D_{\nu}-4d\alpha_{i}D_{\mu}\Box\theta^{\mu\nu}X^{\nu}_{i}+\no\\
    &-& 2(\alpha_{i}\alpha_{j}+\beta_{i}\beta_{j})X^{i}_{\mu} \Box \theta^{\mu\nu}X_{\nu}^{j}+\no\\
    &-& 4(\beta_{i}\beta_{j}+2\beta_{i}\rho_{i}+4\rho_i\rho_j)X^{i}_{\mu}\Box\omega_{\mu\nu}X^{\nu}_{j}
\end{eqnarray}
or
\begin{equation}
        \lag_{K} =(D_{\mu}, X^{i}_{\mu}) \begin{pmatrix} 
            -a\Box\theta^{\mu\nu}&\frac{-c^{t}}{2}\Box\theta^{\mu\nu}\\
            \frac{-c}{2}\Box\theta^{\mu\nu} & -E\Box\theta^{\mu\nu}-T\Box\omega_{\mu\nu}
        \end{pmatrix}\begin{pmatrix}
            D_{\nu}
            \\
            X_{\nu}^{i}
        \end{pmatrix}
\end{equation}
where 
\begin{eqnarray}
    a = 2d^{2};
    c_{i} = (4d\alpha_{2},4d\alpha_{3},\dots, 4d\alpha_{N})
\end{eqnarray}

We can rewrite $\lagK$ as
\begin{equation}
    \lagK = \frac{1}{2}V_{\mu}^{t}K^{\mu\nu}V_{\nu}
\end{equation}
with
\begin{equation}
    V_{\mu} = \begin{pmatrix} 
            D_{\mu}\\X^{i}_{\mu}
            \end{pmatrix}; i =2,...N
\end{equation}
Notice that due to the longitudinal kinetic term, $K_L$, not be necessarily invertible. It must be completed by introducing the gauge-fixing term:

\begin{eqnarray}
    \lagGF &=& \frac{1}{2\xi}[\p_{\mu}(D^{\mu}+\sigma_{i}X^{\mu i})]^{2}\no\\
    &=& \frac{1}{2\xi}[\p_{\mu}(D^{\mu}+\sigma_{i}X^{\mu i})][\p_{\nu}(D^{\nu}+\sigma_{j}X^{\nu j})]
\end{eqnarray}

Defining
\begin{equation}
    \sigma^{t} = (\sigma_{2},\sigma_{3},\dots, \sigma_{N}) 
\end{equation}
and
\begin{equation}
    \Sigma_{(N-1)X(N-1)} = \begin{bmatrix}
    \sigma^{2}_{2}& \dots&\sigma_{2}\sigma_{N}\\
    \vdots&\ddots&\vdots\\
    \sigma_{N}\sigma_{2}&\dots&\sigma_{N}^{2}
    \end{bmatrix}
\end{equation}
one gets,
\begin{equation}
        K^{\mu\nu}_{T} = \begin{pmatrix} 
            -2a&-c^{t}\\
            -c & -2E
        \end{pmatrix}\theta^{\mu \nu}
\end{equation}

\begin{equation}
    K^{\mu\nu}_{L} = \begin{pmatrix} 
            -\frac{1}{\xi}&-\frac{\sigma_{i}^{t}}{\xi}\\
            -\frac{\sigma_{i}}{\Xi} & -2T + \frac{\Sigma_{ij}}{\xi}
        \end{pmatrix}\omega^{\mu \nu}
\end{equation}

\large{ Including the mass term}

\begin{eqnarray}
    \lag_{Massa} &=& \frac{-1}{2}m_{ij}^{2}X^{i}_{\mu}\eta^{\mu\nu}X^{j}_{\nu}\no\\
    &=& \frac{-1}{2}m_{ij}^{2}X^{i}_{\mu}(\theta^{\mu\nu}+\omega^{\mu\nu})X^{j}_{\nu}   
\end{eqnarray}

\begin{equation}
    \lag_{Massa} = \frac{-1}{2}\underbrace{(D_{\mu},X^{i}_{\mu})}_{V_{\mu}^{t}}
    \begin{bmatrix}
    0&0&0&\dots&0\\
    0&m_{22}^{2}&m_{23}^{2}&\dots&m_{2N}^{2}\\
    \vdots&\vdots&\vdots&\ddots&\vdots\\
    0&m_{N2}^{2}&m_{N3}^{2}&\dots&m_{NN}^{2}\\
    \end{bmatrix} \begin{pmatrix} 
            D_{\mu}\\X^{i}_{\mu}
            \end{pmatrix}
\end{equation}

where

\begin{equation} \label{matrix m2}
    M^{2} = -
    \begin{bmatrix}
    0&0&0&\dots&0\\
    0&m_{22}^{2}&m_{23}^{2}&\dots&m_{2N}^{2}\\
    \vdots&\vdots&\vdots&\ddots&\vdots\\
    0&m_{N2}^{2}&m_{N3}^{2}&\dots&m_{NN}^{2}\\
    \end{bmatrix}
\end{equation}

It yields the following propagators equation
\begin{equation}
    \lagK = \frac{1}{2}V_{\mu}^{t}[(K_{T}\Box+M^{2})\theta^{\mu\nu}+[(K_{L}+G_{F})\Box+M^{2})] \omega^{\mu\nu}]V_{\nu}
\end{equation}
where $K_{T}$ and $K_{L}$ symmetry matrices associted to the transverse and longitudinal kinetic contribuitions. $B=K_{L}+G_{F}$ includes the gauge fixing term.

\section{Apendice D. T-fields reparametrization}
${}$\indent

We should establish the physical fields. Consider a parameterization which diagondizes the poles of two point Gree's function. A choice in which the transverse or longitudinal sector is in diagonal form. Here, we choose to diagonalize the kinetic matrix of the transverse sector and then diagonalizes the mass matrix. At Apendice E, we diagonalize the longitudinal sector.

Thus, we opted for the parameterization that completely diagonalizes the transversal sector. Let us denote by $G_{\mu}$ a column vector with the N vector fields $G_{\mu I}$ arranged in this new parameterization. Defining

\begin{eqnarray}
G_{\mu} = \begin{pmatrix}
G_{\mu 1}\\
.\\
.\\
.\\
G_{\mu N}
\end{pmatrix} 
\end{eqnarray}
one relates to the initial parameterization, as follows
\begin{eqnarray}
    G_{\mu} = \Omega^{-1}V_{\mu}, V_{\mu} = \begin{pmatrix}
D_{\mu}\\
X_{\mu}^{i}
\end{pmatrix}
\end{eqnarray}
where the matrix transformation $\Omega$ is given by
\begin{eqnarray}
    &&\Omega=S^{t}\Tilde{K}_{T}^{-1/2}R^{t}\nonumber
    \\
    &&\Omega^{-1}=R\Tilde{K}^{1/2}_{T}S\nonumber
\end{eqnarray}

The matrices that make up $\Omega$ contain the following nature

i) S is the orthogonal matrix that diagonalizes $K_{T}$
\begin{eqnarray}
&&\Tilde{K}_T = SK_{T}S^{t}, (\text{diagonal})
\\
&&K_{T} = S^{t}\Tilde{K}_{T} S
\end{eqnarray}
So in order to avoid the presence of ghosts, in the T-sector, we must have $K_{T}$ positive-defined, that is: $(\Tilde{K}_{T})_{ii}>0 $;

ii) R is the orthogonal matrix that diagonalizes the transformed mass matrix:
\begin{eqnarray}
    &&M^{2} = \Tilde{K}^{-1/2}_{T}S M^{2} S^{t}\Tilde{K}^{-1/2}_{T} = ()^{t}
    \\
    &&m^{2}=RM^{2}R^{t} (\text{diagonal})
    \\
    &&M^{2} = R^{t}m^{2}R
\end{eqnarray}

Notice that $m^{2}_{T}= \Omega^{t}M^{2}\Omega$. The fields $G_{\mu I}$ are combinations of $V_{\mu I}$ and correspond to the poles of the transverse propagators. Thus, in terms of the fields $G_{\mu} = \Omega^{-1}V_{\mu}$ the Lagrangian reads:

\begin{eqnarray}
    \lag = \frac{1}{2}G_{\mu}^{t}[(\Box+m^{2}_{T})\theta^{\mu \nu} + (\Tilde{B}\Box+m^{2}_{T})\omega^{\mu \nu}]G_{\nu}
\end{eqnarray}
where $\Tilde{B}=\Omega^{t}B\Omega=\Omega^{t}(K_{L}+G_{L})\Omega$. The transverse and longitudinal poles are
given respectively by $K^{-1}_{T}$ and $K_{L}^{-1}$.

\subsection{The issue of gauge dependency: at the poles}
\indent${}$
A further study is on the gauge dependence of the poles and the residue of $\Tilde{B}^{-1}m^{2}_{T}$. Since $\Tilde{B}=\Omega^{-1}(K_{L}+G_{F})\Omega$ and $\Tilde{B}$ and $\Omega$ are invertible, then $K_{L }$ and $G_{F}$ are also invertible. 

Considering $\Tilde{B}^{-1} = \Omega^{-1}(K_{L}+G_{F})^{-1}(\Omega^{-1})^{t}$ where you understand $K_{L}$ and $G_{F}$ are the general symmetric matrices:

\begin{eqnarray}
    K_{L} = \begin{pmatrix}
0 & 0\\
0 & s &
\end{pmatrix}; G_{F} = \begin{pmatrix}
\frac{1}{\alpha} & \frac{1}{\alpha}\sigma\\
\frac{1}{\alpha}\sigma^{t} & \frac{1}{\alpha}\sigma \otimes \sigma  
\end{pmatrix}
\end{eqnarray}
it gives,

\begin{eqnarray}
    K_{L}+G_{F} = \begin{pmatrix}
\frac{1}{\alpha} & \frac{1}{\alpha}\sigma\\
\frac{1}{\alpha}\sigma^{t} & s+\frac{1}{\alpha}\sigma \otimes \sigma  
\end{pmatrix} 
\end{eqnarray}
where is a square symmetric matrix $(N-1)\times(N-1)$ that depends on the initial parameters, and $\sigma$ a row vector with elements $\sigma_{i}$. By taking the particular case in which in $\lag_{GF}$ the parameters $\sigma_{i} =0\to \lag_{GF}=\frac{1}{2\alpha}(\partial\cdot D )^{2}$, we have that

\begin{eqnarray}
    &&(K_{L}+G_{F})= \begin{pmatrix}
\frac{1}{\alpha} & 0\\
0 & s
\end{pmatrix}
\\
&&(K_{L}+G_{F})^{-1}= \begin{pmatrix}
\alpha & 0\\
0 & s^{-1}
\end{pmatrix}
\end{eqnarray}

As the propagator poles are given by the eigenvalues of $\Tilde{B}^{-1}m^{2}_{T}$ and that

\begin{eqnarray}
\Tilde{B}^{-1}m^{2}_{T} &&= [\Omega^{-1}(K_{L}+G_{F})^{-1}(\Omega^{-1})^{t}]\Omega^{t}M^{2}\Omega\nonumber
\\
&&=\Omega^{-1}(K_{L}+G_{F})^{-1}M^{2}\Omega,
\end{eqnarray}
one gets that the two sides of this expression are related by a similarity transformation, so the eigenvalues of $\Tilde{B}^{-1}m_{T}^{2}$ are the eigenvalues of $(K_{L}+G_{F })^{-1}M^{2}$. Then just check the eigenvalues of this last matrix:

\begin{eqnarray}
    (K_{L}+G_{F})^{-1} = \begin{pmatrix}
\alpha & 0\\
0 & s^{-1}
\end{pmatrix}
\begin{pmatrix}
0 & 0\\
0 & M^{2}
\end{pmatrix} =
\begin{pmatrix}
0 & 0\\
0 & s^{-1}M^{2}
\end{pmatrix}
\end{eqnarray}
\end{appendix}

This shows us that in the case where $\sigma_{i} = 0$ the non-zero poles of the L-sector are the non-zero eigenvalues of $s^{-1}M^{2}$, therefore completely independent of $\alpha$. However, when analyzing the more general case $\sigma_{i} \neq 0$, we fall back on the result of the previous case. Thus, it is concluded about the independence in the parameters $\alpha, \alpha_{i}$ of $K_{N}$, in the non-zero eigenvalues of the masses of the L-sector.

Thus one can therefore conclude the following regarding the poles of the L-sector:

\begin{enumerate}[(i)]
    \item each zero mass in the T-sector corresponds to a zero mass in the L-sector;
    \item the non-zero masses of the L-sector are non-zero eigenvalues of $s^{-1}M^{2}$, being therefore completely independent of the parameters of $\alpha$ and $\sigma_{i}$;
    \item these masses are independent of the parameterization of the fields.    
\end{enumerate}

Before moving on to the study of the gauge dependence of the residuals at the poles present in the L-sector, we will make a summary of the model's spectroscopy. In the sector of fields $\{V_{\mu}\}$, with fields $D_{\mu}, X_{\mu}^{1},...,X^{N-1}_{\ mu}$ has

\begin{eqnarray}
    &&D_{\mu} \to \text{null mass};
    \\
    &&X^{i}_{\mu}  \to \left\{N-1 spin-1, N-1 spin-0\right\} \text{basically massive}
\end{eqnarray}

In terms of fields set $\{G_{\mu}\}$, the fields $G_{\mu I}$ contain:

\begin{enumerate}[(i)]
    \item \textbf{in the $T$ sector} we read N poles corresponding to the masses of the N spin-1 fields we have a null mass and N-1 masses in general;
    \item \textbf{in the sector-$L$} 1 null pole (corresponding to the compensating spin-0 of the photon) and N-1 "gauge-independent" poles corresponding to the spin-0 fields.
\end{enumerate}

It should be noted that the massive fields are not Proca fields, as they contain the longitudinal part; and that if the eigenvalues of $s^{-1}M^{2}$ are negative, they will correspond to tachyons.

\subsection{Poles and residues at sector-L}
\indent${}$

Consider to the analysis of the L-sector residues, 

\begin{eqnarray}
    <G_{\mu}G_{\nu}>_{L} = \frac{1}{\Box+\Tilde{B}^{-1}m^{2}_{T}}\Tilde{B}^{-1}P^{L}_{\mu \nu}
\end{eqnarray}
where $\Tilde{B}=\Omega^{t}(K_{L}+G_{F})\Omega; \text{ }m^{2}_{T} = \Omega^{t}M^{2}\Omega$;

\begin{eqnarray}
    K_{L} = \begin{pmatrix}
0 & 0 & \dots & 0\\
0 & s \\
\\
0
\end{pmatrix}; G_{F} = \begin{pmatrix}
\frac{1}{\alpha} & \frac{1}{\alpha}\sigma\\
\frac{\sigma^{t}}{\alpha} & \frac{1}{\alpha}
\end{pmatrix} ; M^{2} = \begin{pmatrix}
0 & 0 & \dots & 0\\
0 & M^{2} \\
\\
0
\end{pmatrix} 
\end{eqnarray}
then,
\begin{eqnarray}
    \Tilde{B}^{-1}m^{2}_{T} = \Omega^{-1}(K_{L}+G_{F})^{-1}M^{2}\Omega
\end{eqnarray}
which gives
\begin{eqnarray}
    <G_{\mu} G_{\nu}>_{L} = \frac{1}{\Box + \Omega^{-1}(K_{L}+G_{F})^{-1}M^{2}\Omega}\Omega^{-1}(K_{L}+G_{F})^{-1}(\Omega^{-1})^{t}\omega_{\mu \nu}\nonumber\\
\end{eqnarray}
Then, after expanding the denominator, conveniently, it gives:

\begin{eqnarray}
    <G_{\mu}G_{\nu}>_{L} = \Omega^{-1}\left(\frac{1}{\Box+(K_{L}+G_{F})^{-1}M^{2}}(K_{L}+G_{F})\right)(\Omega^{-1})^{t}\omega_{\mu \nu}
\end{eqnarray}
where, after some algebraism, it turns out that
\begin{eqnarray}
    \left[\frac{1}{\Box+(K_{L}+G_{F})^{-1}M^{2}}\right](K_{L}+G_{F})^{-1}=\begin{pmatrix}
u & v \\
v^{t} & t
\end{pmatrix}
\end{eqnarray}
where

\begin{eqnarray}
    &&u = \frac{1}{\Box}\{\alpha\left[1-\frac{\sigma}{\alpha}(s+\frac{1}{\alpha}\sigma^{t}\sigma)^{-1}\sigma^{t}\right]^{-1}  - \sigma s^{-1}M^{2}(\Box+s^{-1}M^{2})^{-1}\cdot\nonumber
    \\
    &&\cdot (s+\frac{\sigma^{t}\sigma}{\alpha})^{-1}\sigma^{t}\left[1-\frac{\sigma}{\alpha}(s+\frac{1}{\alpha}\sigma^{t}\sigma)^{-1}\sigma^{t}\right]^{-1}\}
    \\
    &&v= -\frac{1}{\Box}\sigma s^{-1}[1-M^{2}s^{-1}(\Box+M^{2}s^{-1})^{-1}];
    \\
    &&t=(\Box+s^{-1}M^{2})^{-1}s^{-1}
\end{eqnarray}

Concluding,
\begin{eqnarray}
    <G_{\mu} G_{\nu}>_{L} = \Omega^{-1}\begin{pmatrix}
u & v\\
v^{t} & t
\end{pmatrix}(\Omega^{-1})^{t}\omega_{\mu \nu}
\end{eqnarray}
Notice that

\begin{eqnarray}
    &&u=u(\alpha; \sigma_{i});
    \\
    &&v=v(\sigma_{i});
    \\
    && \text{does not depend on the parameters } \alpha \text{ and } \sigma_{i}
\end{eqnarray}

Let's take: $\Omega^{-1} =  \begin{pmatrix}
x & Y\\
W & Z
\end{pmatrix}$ where:

x is a number $=\Omega^{-1}$;

Y is a row matrix $1\times(N-1)$;

W is a column matrix $(N-1)\times 1$;

Z is a matrix $(N-1)\times(N-1)$;  

It gives,
\begin{eqnarray}
    <G_{\mu}G_{\nu}>_{L} = \begin{pmatrix}
ux^2 + x(vY^{t}+Yv^{t})+utY^{t} & uxW^{t}+xvZ^{t}+Yv^{t}Wt+YtZ^{t} \\
uxW+xZv^{t}+WvY^{t}+ZtY^{t} & uWW^{t}+WvZ^{t}+Zv^{t}W^{t}+ZtZ^{t}
\end{pmatrix}\cdot \omega_{\mu \nu}\nonumber
\\
\end{eqnarray}

Our first question is to identify which are the propagators $<G_{\mu}G_{\nu}>_{L}$ whose residues at the poles bring dependence on $\alpha$. Notice that the only dependency on $\alpha$ comes through $u$. It Yield,

 \begin{eqnarray}
     (i) <G_{\mu 1}G_{\nu 1}>_{L} = x^{2}u+x(vY^{t}+Yv^{t})+YtY^{t};
 \end{eqnarray}

Eq. (D.30) dependence on $\alpha$ is found in the term $x^{2}u(k^{2};\alpha;\sigma_{i})$. On the other hand, for whatever the pole ($k^{2}=0 \text{ or } k^{2}$ a certain eigenvalue of $s^{-1}M^{2}$) the dependency in $\alpha$ presented by the residue $u(k^{2}; \alpha;\sigma_{i})$ at the pole in question does not disappear. Therefore, the propagator $<G_{\mu 1}G_{\nu 1}>_{L}$ will have residues at the poles $k^{2} = 0$ or $k^{2}=\mu \neq 0$ with explicit dependence on $\alpha \neq 0$, that is, it is the element $(\Omega^{-1})_{11} = x$ who controls the dependence (or not) on $\alpha$ of the residues presented by the propagator $<G_{\mu 1}G_{\nu 1}>_{L}$ at its pole.

Considering the mixing propagators of the type $<G_{\mu 1}G_{\nu I}>_{L}$ with $I \geq 2$. These are given by

 \begin{eqnarray}
     <G_{\mu 1}G_{\nu I}>_L = uxW^{t} + xvZ^{t} + yv^{t}W^{t} + YtZ^{t}
 \end{eqnarray}
As we have previously analyzed, whatever the pole $(k^{2} = 0 \text{ or } k^{2} =S^{-1}M^{2})$, the residue of $a(k^{2 }, \alpha, \sigma)$ has explicit dependency on $\alpha$. Therefore, the only way to cancel the dependence on $\alpha$ is through the elements $x$ or $W_{i}$. If $x \neq 0$ remains the possibility that some $W_{i} =0$ e, then the propagator $<G_{\mu 1}G_{\nu I}>_{L}$ has no dependence on $\alpha$.

Considering the propagators $<G_{\mu I}G_{\nu J}>_{L}$ with $I,J \neq Z$, it gives

\begin{eqnarray}
    <G_{\mu I}G_{\nu J}>_{L} = uWW^{t}+Wvz^{t}+Zv^{t}W^{t}+Zv^{t}W^{t}+ZtZ^{t}
\end{eqnarray}

At this case the $\alpha-$ dependency is canceled on all these propagators.

\section{Apendice E. L-sector reparametrization}
\indent${}$
Starting from initial $\lag$

\begin{eqnarray}
    \lag = \frac{1}{2}V_{\mu}^{t}(\Box K_{T} +M^{2})P^{\mu \nu}_{T} + \frac{1}{2}V_{\mu}^{t}(\Box B+M^{2})P^{\mu \nu}_{L}V_{\nu}
\end{eqnarray}
where $B=K_{L}+G_{F}$, we introduce the diagonalization of the L-sector

\begin{eqnarray}
    V_{\mu} = \Omega_{L}L_{\mu}
\end{eqnarray}
where $L_{\mu}$ are the physical fields that diagonalize the L-sector. Considering

\begin{eqnarray}
    \Omega_{L}B\Omega_{L} = \mathbf{1} \text{ and } \Omega^{t}_{L}M^{2}\Omega_{L} = m_{L}^{2}
\end{eqnarray}
we get

\begin{eqnarray}
    \Omega_{L}M^{2}\Omega_{L} = \Omega_{L}^{-1}(B^{-1}M^{2}) = m_{L}^{2}
\end{eqnarray}
that is, the physical masses of the L-sector that appear as the diagonal elements of $m_{L}^{2}$ are the eigenvalues of the matrix $(B^{-1}M^{2})$ .

At this way, we will introduce in the T-sector the transformation $\Omega_{L}$ that carried out the passage in the fields $L_{\mu}$ of the L sector:

\begin{eqnarray}
    \lag = \frac{1}{2}L_{\mu}^{t}(\Box \Tilde{K}_{T}+\Tilde{M}^{2})P^{\mu \nu}_{T}L_{\nu}+\frac{1}{2}L_{\mu}^{t}(\Box +m^{2}_{L})P^{\mu \nu}_{L}L_{\nu}
\end{eqnarray}
where $\Tilde{K}_{T}=\Omega^{t}_{L}K_{T}\Omega_{L}, \Tilde{M}^{2} = \Omega^{t}_{ L}M^{2}\Omega_{L}$. Notice, then, that $\Tilde{K}_{T}$ is not necessarily positive since the transformation $\Omega_{L}\Omega_{T}\Omega_{L}$ is not of similarity $(\Omega^ {t}_{L}\neq \Omega^{-1}_{L})$ such that $K_{T}$ and $\Tilde{K}_{T}$ do not have the same eigenvalues. However, it is interesting to observe that

\begin{eqnarray}
    \Tilde{K}^{-1}_{T}M^{2} = (\Omega^{-1}_{L}K^{-1}_{T}\Omega^{-1 t}_{L})(\Omega^{t}_{L}M^{2}\Omega^{-1 t}_{L}) = \Omega^{-1}_{L}(K^{-1}_{T}M^{2})\Omega_{L}
\end{eqnarray}
which shows us that the poles of the propagators $<L_{\mu}L_{\nu}>_{L}$ are the eigenvalues of $K^{-1}_{T}M^{2}$ which, in turn, are the diagonal elements of the matrix $m^{2}_{T}$ of the physical masses at sector-T. 

\section{Apendice F. $V_{\mu}$-referencial}
\indent${}$

Considerando a noção de diversidade para um conjunto de campos $\{A_{\mu I}\}$ desenvolveu-se no capítulo anterior a chamada Lagrangeana abeliana não-linear, eq. (2.1). O propósito aqui estará em entendê-la em termos de sistema de referência de campos. De posse que a teoria oferece aosseus números quânticos (quanta $\equiv$ massa, spin, carga, CPT).

Desta, maneira primeiramente estudaremos a formsa matricial para N-campos vetoriais $V_{\mu}^{t} \equiv (A_{\mu 1}...A_{\mu N})$ do seu setor livre $L_{0}$ composto pelos termos cinéticos, massa e gauge-fixing.

\begin{eqnarray}
    \lag_{0} = \frac{1}{2}V_{\mu}^{t}[(K_{T}\Box+M^2)P^{\mu \nu}_{T} + ((K_{L}+G_{F})\Box+M^{2})\omega^{\mu \nu}]V_{\nu}
\end{eqnarray}
onde a condição de realidade nos leva a 

\begin{eqnarray}
K_{T}=K_{T}^{t}, \text{ } K_{L} = K_{L}^{t}, \text{ }M^{2}=(M^{2})^{t}, \text{ }G_{F} = G_{F}^{t}  
\end{eqnarray}
e a dizer que essas matrizes-$N\times N$ simétricas e reais são passíveis de serem diagonalizadas por transformações ortpgonais.

Há duas opções de diagonalização: ou para o setor transverso ou para o setor longitudinal. Procedendo-se à digonalização: ou para o setor transversa, podemos extrair os resultados a partir do que se fez para o caso de Klein-Gordon estendido, onde $\phi \equiv V_{\mu}$  e $\Phi \equiv G_{\nu}$
\begin{eqnarray}
   V_{\mu} \Omega G_{\mu}, & G_{\mu} = \Omega^{-1}V_{\mu}
\end{eqnarray} 
onde $\Omega = S^{t}\Tilde{K}^{1/2}R^{t}$, entendendo-se que S diagonaliza a matriz cinética transversal, $\Tilde{K}^{1/2}$ normaliza os auto-valores da diagonal com a identidade e R diagonaliza a matriz de massa resultante dessas transformações. Assim relembre que $\Omega\Omega^{-1} = \mathbf{1}$ mas não é em geral unitária $\Omega^{\dagger}\Omega = S^{\dagger}\tilde{K}S \neq \mathbf{1}$ (a não ser que a matriz cinética seja diagonal).

Rewriting this bilinear sector in terms of fields $G_{\mu}$, we obtain that:

\begin{eqnarray}
    \lag_{0} = \frac{1}{2}G_{\mu}(\Box+m^{2}_{T})\theta^{\mu \nu}G_{\nu} + \frac{1}{2}G_{\mu}^{t}(\Tilde{K}_{L}+m^{2}_{T})\omega^{\mu \nu}G_{\nu}
\end{eqnarray}
where
\begin{eqnarray}
    \Omega^{t}K\Omega^{t} = \mathbf{1}, & m^{2} = \Omega^{t}M^{2}\Omega, & \Tilde{K}_{L} = \Omega^{t}\Tilde{K}_{L}\Omega
\end{eqnarray}
and for the sake of simplicity we will not include the gauge-fixing matrix in the following studies.

For the sake of consistency, we will unify the above results.

\begin{eqnarray}
    &&\lag_0 = \frac{1}{2}V_{\mu}S^{t}SK_{T}S^t S\Box \theta^{\mu \nu}V_{\nu}+\frac{1}{2}V_{\mu}S^{t}SK_{L}S^t S\Box\omega^{\mu \nu}V_{\nu} \nonumber
    \\
    &&+\frac{1}{2}V_{\mu}S^{t}SM^{2}S^{t}SV^{\mu}\nonumber
    \\
    &&=\frac{1}{2}\left[\Tilde{V}_{\mu}^{t}\Tilde{K}_{T}\Box\theta^{\mu \nu}\Tilde{V}_{\nu}+\Tilde{V}_{\mu}^{t}\Tilde{K}_{L}\Box\omega^{\mu \nu}V_{\nu}+\tilde{V}_{\mu}\Tilde{M^{2}}V^{\mu}\right]
\end{eqnarray}
where $\Tilde{V}_{\mu} = SV_{\mu}$, $\Tilde{K}=SKS^{t}$. Redefining

\begin{eqnarray}
    \Tilde{K} = k_{i}\delta_{ij},& \sqrt{k_{i}}\Tilde{V}_{\nu} = \Tilde{\Tilde{V_{\nu}}}
\end{eqnarray}

\begin{eqnarray}
    \lag_{0} = \left[\Tilde{\Tilde{V_{\mu}^{t}}}\Box\theta^{\mu \nu}\Tilde{\Tilde{V_{\nu}}}+\Tilde{\Tilde{V^{t}_{\mu}}}\Tilde{\Tilde{K_{L}}}\Box\omega^{\mu \nu}\Tilde{\Tilde{V_{\nu}}}+ \Tilde{\Tilde{V^{t}_{\mu}}}\Tilde{\Tilde{M}}\Tilde{\Tilde{V^{\mu}}}\right]
\end{eqnarray}
where $\Tilde{\Tilde{K}}_{L} = \Tilde{K}^{-1/2}\Tilde{K}_{L}\Tilde{K}^{-1/2}$, $\Tilde{\Tilde{M}} = \Tilde{K}^{-1/2}\Tilde{M}\Tilde{K}^{-1/2}$. With $\Tilde{\Tilde{M^{2}}} = (\Tilde{\Tilde{M^{2}}})^{t}$, this matrix can be diagonalized. Let R be the diagonal matrix that diagonalizes $\Tilde{\Tilde{M^{2}}}$.

\begin{eqnarray}
    m^{2} = R\Tilde{\Tilde{M^{2}}}R^{t}, & G_{\mu} = R\Tilde{\Tilde{V}}_{m}u
\end{eqnarray}
So, these results lead us to

\begin{eqnarray}
   && G_{\mu} = (R\Tilde{K}^{1/2}S)V_{\mu} = \Omega^{-1}V_{\mu}\nonumber
   \\
   &&\Tilde{K}_{L} = (R\Tilde{K}^{-1/2}S)K_{L}(S^{t}\Tilde{K}^{-1/2}R^{t})\nonumber
   \\
   &&m^{2} = (R\Tilde{K}^{-1/2}S)M^{2}(S^{t}\Tilde{K}^{-1/2}R^{t})
\end{eqnarray}
that reproduce the expressions () and ()

The next step is to study $\lag_{int}$

\begin{eqnarray}
    \lag_{int} = \lambda_{ijk}\partial_{\mu}V_{\nu}^{i}V^{\mu j}V^{\nu k} + s_{ijk}\partial_{\mu}V^{\mu i}V_{\nu}^{j}V^{\nu k} + \Lambda_{ijkl}V^{i}_{\mu}V^{j}_{\nu}V^{\mu k}V^{\nu l}
\end{eqnarray}

that in matrix form

\begin{eqnarray}
    \lag_{int} = \partial_{\mu}V_{\nu}[V_{\mu}^{t}\lambda V_{\nu}] + \partial_{\mu}V^{\mu t}[V_{\nu}^{t}sV^{\nu}] + V_{\mu}^{t}V_{\nu}^{t}\Lambda V^{\mu}V^{\nu}
\end{eqnarray}

Using $V_{\nu} = \Omega G_{\mu}$

\begin{eqnarray}
    \lag_{int} = \partial_{\mu}G_{\nu}][G^{\mu}\Tilde{\lambda}G^{\nu}] + \partial_{\mu}G^{\mu}[G_{\nu}\tilde{s}G^{\nu}] + G_{\mu}^{t}[G_{\nu}\Tilde{\Lambda}G^{\mu}]G^{\nu}
\end{eqnarray}
where
\begin{eqnarray}
&&\tilde{\lambda} = \Omega^{t}\Omega^{t}\lambda\Omega\nonumber
\\
&&\tilde{s} = \Omega^{t}\Omega^{t}s\Omega
\\
&&\tilde{\Lambda} = \Omega^{t}\Omega^{t}\Lambda\Omega
\end{eqnarray}

Again, we will rephrase the Lagrangian interaction terms in explicit field terms. Let's start with the initial Lagrangian itself

\begin{eqnarray}
    \lag_{int} &&= (\partial_{\mu}V^{I}_{\nu})\left[V^{\mu J}(\lambda_{I})_{JK}V^{\nu K}\right] + (\partial_{\mu}V^{\mu I})\left[V_{\nu}^{J}(s_{I})_{JK}V^{\nu K}\right] +\nonumber
    \\
    &&+V_{\mu}^{I}\left[V_{\nu}^{I}(\Lambda_{IL})_{JK}\right]V^{\nu L}
\end{eqnarray}

Considering the rotation again

\begin{eqnarray}
    &&V_{\mu} = s^{t}\tilde{V}_{\mu} \to V_{\mu I} = S_{MI}\tilde{V}_{\mu M}\nonumber
    \\
    &&V_{\mu} = s^{t}\tilde{V}_{\mu}^{t} \to \tilde{V}^{t}_{\mu I} = \tilde{V}_{\mu M}S_{MI}
\end{eqnarray}

and that $S_{MI}S_{IN} = \delta_{MN}$ $(S^{t}S = 1)$, we have as an example for the trilinear term

\begin{eqnarray}
    &&V^{\mu t}_{J}(\lambda_{I})_{JK}V_{K}^{\nu} = V^{\mu t}_{J}\delta_{JL}(\lambda_{I})_{L \theta}V^{\mu}_{K}\nonumber
    \\
    &&=V^{\mu t}_{J}S_{JN}\left[S_{NL}(\lambda_{I})_{L \theta}S_{\theta P}\right]S_{PK}V^{\nu}_{K}\nonumber
    \\
    &&\tilde{V}^{\mu t}_{N}[\tilde{\lambda}_{I}]\tilde{V}^{\nu}_{P} \textbf{, } (\Tilde{\lambda}_{I})_{NP} = S_{NL}(\lambda_{I})_{L \theta}S_{\theta P} 
\end{eqnarray}

Completing the trilinear time in the fields and placing the delta conveniently:

\begin{eqnarray}
    \partial_{\mu}V^{I}_{\nu} \delta_{IL}[\tilde{V}^{\mu}_{N}(\tilde{\lambda_{I}})_{NP}\tilde{V}^{\nu}_{P}] = (\partial_{\mu}\tilde{V}_{\theta})S_{\theta L}\left[\tilde{V}^{\mu}_{N}(\tilde{\lambda_{L}})_{NP}\tilde{V}^{\nu}_{P}\right]
\end{eqnarray}
Continuing on, $\tilde{\tilde{V}}_{\mu} = \tilde{K}^{1/2}\tilde{V}_{\mu}$ which in terms of the peasant fields means:

\begin{eqnarray}
    \tilde{\tilde{V}}_{\mu I} = (k_{I})^{1/2}\tilde{V}_{\mu I} \to  \tilde{V}_{\mu I} = (k_{I})^{1/2} \tilde{\tilde{V}}_{\mu I}
\end{eqnarray}
So adding $\delta_{\theta M} = k_{\theta N}^{1/2}K_{N M}^{-1/2}$, we get:

\begin{eqnarray}
    &&(\partial_{\mu}\tilde{V}_{\theta})\delta_{\theta M}S_{ML}[\tilde{V}^{\mu}_{N}\delta_{NR}\tilde{\lambda}_{LRQ}\delta_{QP}\tilde{V}^{\nu}_{P}]\nonumber
    \\
    = \partial_{\mu}\tilde{V}_{\theta}\tilde{k}^{1/2}_{ND}\tilde{k}^{-1/2}_{DR}(\tilde{L})_{QR}\tilde{k}^{-1/2}_{QX}\tilde{k}^{1/2}_{XP}\tilde{V}^{\nu}_{P}
\end{eqnarray}
which leads us to the result

\begin{eqnarray}
    \partial_{\mu}\tilde{\tilde{V}}_{\nu \theta}\tilde{k}^{-1/2}_{\theta M}S_{ML}[\tilde{\tilde{V}}^{\mu}_{N}(\tilde{\lambda}_{L})_{NP}\tilde{\tilde{V}}^{\nu}_{P}]
\end{eqnarray}
where $(\tilde{\tilde{\lambda}}_{L})_{NP} = \tilde{k}^{-1/2}_{NR}(\tilde{\lambda}_{L})_{RQ}\tilde{k}^{-1/2}_{QP}$
Finally, rotating this term with the orthogonal matrix R that diagonalizes the mass matrix, results in terms of component fields $G_{\mu I} = R_{IJ}\tilde{\Tilde{V}}_{\mu J }$.

Adding the $\delta_{IJ} = R_{IK}R_{KJ}$, we have

\begin{eqnarray}
&&\partial_{\mu}\tilde{\Tilde{V}}_{\nu \theta}\tilde{k}^{-1/2}_{KM}S_{ML}\left[\tilde{\tilde{V}}^{\mu}_{N}\delta_{NT}(\tilde{\tilde{\lambda_{L}}})_{TQ}\delta_{QP}\tilde{\tilde{V}}^{\nu}_{P}\right]\nonumber
\\
&&=(\partial_{\mu}G_{\nu S})R_{SR}\tilde{k}^{-1/2}_{RM}S_{ML}[G_{N}^{\mu}(\Bar{\lambda}_{L})_{NP}G_{P}^{\nu}]\nonumber
\\
&&=(\partial_{\mu}G_{\nu S})\Omega_{SL}^{-1}\left[G_{N}^{\mu}(\Bar{\lambda_{L}})_{NP}G^{\nu}_{P}\right]
\end{eqnarray}
where $(\Bar{\lambda_{L}})_{NP} = R_{NT}(\tilde{\tilde{\lambda}}_{L})_{TQ}R_{PQ}$

By analogy, we can extend to the other terms such that the interaction Lagrangian written in terms of the component fields is written:

\begin{eqnarray}
    \lag_{int} &&= (\partial_{\mu}G_{\nu})_{I}\Omega_{JI}\left[G_{L}^{\mu}(\Bar{\lambda}_{J})_{LM}G^{\nu}_{M}\right] +  (\partial_{\mu}G_{\mu})_{I}\Omega_{JI}\left[G_{\nu L}(\Bar{s}_{J})_{LM}G^{\nu}_{M}\right]\nonumber
    \\
    &&+G_{\mu I}\Omega_{JI}\left[G_{\nu K}(\Bar{\Lambda}_{JK})_{KM}G_{M}^{\nu}\right]\Omega_{LN}G^{\mu}_{N}
\end{eqnarray}
where in terms of the array elements

\begin{eqnarray}
    &&(\Bar{\lambda}_{J})_{LM} = \Omega_{TL}(\lambda_{J})_{TS}\Omega_{SM}\nonumber
    \\
    &&(\Bar{s}_{J})_{LM} = \Omega_{TL}(s_{J})\Omega_{SM}\nonumber
    \\
    &&(\Bar{\Lambda}_{JL})_{KM} = \Omega_{TK}(\Lambda_{JL})_{TS}\Omega_{SM}
\end{eqnarray}

Implementing the notation:

\begin{eqnarray}
    &&\Omega_{JI}(\Bar{\lambda}_{J})_{LM} = (\Bar{\Bar{\lambda}})_{LM}, \text{ } \Omega_{JI}(\Bar{s}_{J})_{LM} = (\Bar{\Bar{s}}_{I})_{LM}\nonumber
    \\
    &&\text{e }\Omega_{JI}(\Bar{\Lambda}_{JL})_{LN} = (\Bar{\Bar{\Lambda}}_{IN})_{KM}
\end{eqnarray}
we will have
\begin{eqnarray}
    \lag_{int} &&= (\partial_{\mu}G_{\nu I})\left[G_{L}^{\mu}(\Bar{\Bar{\lambda}}_{I})_{LM}G^{\nu}_{M}\right] + (\partial_{\mu}G^{\mu}_{I})\left[G_{L}^{\mu}(\Bar{\Bar{s}}_{I})_{LM}G_{M}^{\nu}\right]\nonumber
    \\
    &&G_{\mu I}\left[G_{\nu K}(\Lambda_{IN})_{KM}G^{\nu}_{M}\right]G^{\mu}_{N}
\end{eqnarray}
where

\begin{eqnarray}
    &&\Bar{\Bar{\lambda}}_{ILM} = \Omega_{JI}\Omega_{TL}\lambda_{JTS}\Omega_{SM}\nonumber
    \\
    &&\Bar{\Bar{s}}_{ILM} = \Omega_{JI}\Omega_{TL}s_{JTS}\Omega_{SM}\nonumber
    \\
    &&\Bar{\Bar{\Lambda}}_{INKM} = \Omega_{JI}\Omega_{TL}\Lambda_{JLTS}\Omega_{SM}\Omega_{LN}
\end{eqnarray}

\section{Apendice G. Referencial $V_{\mu}$ and CPT invariance discrete symmetries}
\indent${}$

\begin{eqnarray}
    L &&= \frac{1}{2}V_{\mu}^{t}[(K\Box+M^{2})\theta^{\mu \nu}_{T}+((K+B)\Box+M^{2})\omega^{\mu \nu}_{L}]V_{\nu}\nonumber
    \\
    &&\partial_{\mu}V_{\nu}^{t}\lambda V^{\mu}V^{\nu} + V_{\mu}^{t}V_{\nu}^{t}\Lambda V^{\mu}V^{\nu}
\end{eqnarray}
with the following base change conditions:

\begin{eqnarray}
    V_{\mu} = \Omega G_{\mu}
\end{eqnarray}
and
\begin{eqnarray}
    \Tilde{B} = \Omega B \Omega, & \Tilde{\lambda} = \Omega^{t} \lambda \Omega \Omega & \Tilde{\Lambda} = \Omega^{t}\Omega^{t}\Lambda\Omega \Omega 
\end{eqnarray}
Given that
\begin{eqnarray}
    V_{\mu}(x) \stackrel{PCT}{\to} U_{TCP}V_{\mu}(x)U^{-1}_{TCP} = V'_{\mu}(x') = \mathcal{N}V_{\mu}(x)
\end{eqnarray}
one definies $\mathcal{N} = \Omega \eta_{PCT}\Omega^{-1}$

In this way we will study separately each term in the reference-$V_{\mu}$:

(i) transverse part

\begin{eqnarray}
    V^{t}(K\Box+M^{2})\theta_{T}V \stackrel{PCT}{\to} V^{t}\mathcal{N}^{t}(K\Box+M^{2})\theta\mathcal{N}V
\end{eqnarray}
Considering
\begin{eqnarray}
    \mathcal{N}^{t}K\mathcal{N} && = (\Omega^{-1})^{t}\eta_{PCT}\Omega^{t}K\Omega \eta_{PCT}\Omega^{-1} = (\Omega^{-1})^{t}\Omega^{-1} = K\nonumber
    \\
    \mathcal{N}^{t} M^{2} \mathcal{N} && = (\Omega^{-1})^{t}\eta_{PCT}\Omega^{t}M^{2}\Omega \eta_{PCT}\Omega^{-1}
    \\
    &&= (\Omega^{-1})^{t}\eta_{PCT}\Omega^{t}m^{2}\Omega \eta_{PCT}\Omega^{-1} = M^{2}
\end{eqnarray}
which means that the transverse term is automatically PCT-invariant.

(ii) Longitudinal part

\begin{eqnarray}
    V^{t}((K+B)\Box + M^{2})\omega V \stackrel{PCT}{\to} V^{t}\mathcal{N}^{t}((K+B)\Box + M^{2})\mathcal{N}\omega V
\end{eqnarray}

As $\mathcal{N}^{t}K\mathcal{N} = K$, $\mathcal{N}^{t}M^{2}\mathcal{N} = M^{2}$, it remains to understood the relationship

\begin{eqnarray}
    \mathcal{N}^{t}B\mathcal{N} = (\Omega^{-1})^{t}B \textbf{ }\Omega\eta_{PCT}\Omega^{-1}
\end{eqnarray}

Considering that $\Omega B \Omega = \Tilde{B}$ and $\eta_{PCT}\Tilde{B}\eta_{PCT} = B$ (due to invariance in the reference-$G_{\mu}$), it follows that:

\begin{eqnarray}
    \mathcal{N}^{t}B\mathcal{N} = (\Omega^{-1})\Tilde{B}\Omega^{-1} = B
\end{eqnarray}
that is, the longitudinal term is invariant PCT but having as a condition the relation $\eta_{PCT}\Tilde{B}\eta_{PCT} = \Tilde{B}$ coming from the imposition of invariance in the reference-$G_{\mu}$.
\\
(iii) Trilinear term

Using $\partial'_{\mu} = - \partial_{\mu}$:

\begin{eqnarray}
    &&\partial_{\mu}V_{\nu}^{t}\lambda V^{\mu}V^{\nu} \stackrel{PCT}{\to} - \partial_{\mu}V_{\nu}^{t}\mathcal{N}^{t}\lambda \mathcal{N}\mathcal{N}V^{\mu}V^{\nu}
    \\
    &&\mathcal{N}^{t}\lambda\mathcal{N}\mathcal{N} = (\Omega^{-1})^{t}\eta_{PCT}\Omega^{t}\lambda \Omega \eta_{PCT}\Omega^{-1}\Omega \eta \Omega^{-1}
\end{eqnarray}

Using $\Omega^{t}\lambda\Omega \Omega = \Tilde{\lambda}$

\begin{eqnarray}
    \mathcal{N}^{t}\lambda\mathcal{N}\mathcal{N} = (\Omega^{-1})^{t}\eta_{PCT}\Tilde{\lambda}\eta_{PCT} \eta_{PCT} \Omega^{-1}\Omega^{-1}
\end{eqnarray}

Using the invariance condition in the frame-$G_{\mu}$, $\eta_{PCT}\Tilde{\lambda}\eta_{PCT}\eta_{PCT} = - \Tilde{\lambda}$

\begin{eqnarray}
    \mathcal{N}^{t}\lambda\mathcal{N}\mathcal{N} = - (\Omega^{-1})^{t}\eta_{PCT}\Tilde{\lambda}\eta_{PCT}\eta_{PCT}\Omega^{-1}\Omega^{-1}
\end{eqnarray}

Using the invariance condition in the frame-$G_{\mu}$, $\eta_{PCT}\Tilde{\lambda}\eta_{PCT}\eta_{PCT} = - \tilde{\lambda}$, one gets

\begin{eqnarray}
    \mathcal{N}^{t}\lambda\mathcal{N}\mathcal{N} = - (\Omega^{-1})^{t}\Tilde{\lambda}\Omega^{-1}\Omega^{-1} = -\lambda
\end{eqnarray}
i.e.

\begin{eqnarray}
    \partial_{\mu}V_{\nu}^{t}\lambda V^{\mu}V^{\nu} \stackrel{PCT}{\to} \partial_{\mu}V_{\nu}^{t}\lambda V^{\mu}V^{\nu}
\end{eqnarray}

(iv) Quadrilinear term

\begin{eqnarray}
    V_{\mu}^{t}V_{\nu}^{t}\lambda V^{\mu}V^{\nu} \stackrel{PCT}{\to} V_{\mu}^{t}V_{\nu}^{t}\mathcal{N}^{t}\mathcal{N}^{t}\Lambda\mathcal{N}\mathcal{N}V^{\mu}V^{\nu}
\end{eqnarray}

\begin{eqnarray}
    \mathcal{N}^{t}\mathcal{N}^{t}\Lambda \mathcal{N} \mathcal{N} &&= \left((\Omega^{-1})^{t}\eta_{PCT}\Omega^{t}\right)_{i}\left((\Omega^{-1})^{t}\eta_{PCT}\Omega^{t}\right)_{j}\cdot \nonumber
    \\
    &&\cdot \Lambda_{ijkl}(\Omega \eta_{PCT}\Omega^{-1})_{\kappa}(\Omega \eta_{PCT}\Omega^{-1})_{l}
\end{eqnarray}

Using $\Omega^{t}\Omega^{t}\Lambda \Omega \Omega = \Tilde{\Lambda}$:

\begin{eqnarray}
\mathcal{N}^{t}\mathcal{N}^{t}\Lambda\mathcal{N}\mathcal{N} = (\Omega^{-1})^{t}(\Omega^{-1})^{t}\eta_{PCT}\eta_{PCT}\Tilde{\Lambda}\eta_{PCT}\eta_{PCT}\Omega^{-1}\Omega^{-1}
\end{eqnarray}
By invariance in the $G_{\mu}$-framework, we have $\eta_{PCT}\eta_{PCT}\Tilde{\Lambda}\eta_{PCT}\eta_{PCT} = \Tilde{\Lambda}$. It yields,

\begin{eqnarray}
    \mathcal{N}^{t}\mathcal{N}^{t}\Lambda\mathcal{N}\mathcal{N} = (\Omega^{-1})^{t}(\Omega^{-1})^{t}\Tilde{\Lambda}\Omega\Omega = \Lambda
\end{eqnarray}
Then
\begin{eqnarray}
    V^{t}V^{t}\Lambda VV \stackrel{PCT}{\to} V^{t}V^{t}\Lambda VV 
\end{eqnarray}

Conclusion: the invariance conditions of the $G_{\mu}$-frame ensure invariance in the $V_{\mu}$ frame, however, the PCT invariance in $G_{\mu}$ is not immediate.

\end{document}